\documentclass[%
aip,
amsmath,amssymb,
floatfix,
]{revtex4-1}

\usepackage{graphicx}
\usepackage{placeins}
\usepackage{dcolumn}
\usepackage{bm}
\usepackage[utf8]{inputenc}
\usepackage[T1]{fontenc}
\usepackage{mathptmx}
\usepackage{xcolor}
\begin{document}
	\title[]{A numerical study of gravity-driven instability in strongly coupled dusty plasma. Part 2: Hetero-interactions between a rising bubble and a falling droplet}
	
	\author{Vikram S. Dharodi}%
	\email{dharodiv@msu.edu}
	\affiliation{Mechanical Engineering, Michigan State University, East Lansing, Michigan 48824, USA.}
  \affiliation{Institute for Plasma Research, HBNI, Bhat, Gandhinagar 382 428, India}\thanks{Current affiliation}
	\date{\today}
	\begin{abstract}
In part I \cite{dharodi2020numerical} (accepted in JPP), we simulated individual dynamics of a rising bubble and a falling droplet in an incompressible limit of a strongly coupled dusty plasma (SCDP) using a generalized hydrodynamic viscoelastic fluid model. In interest to understand hetero- (bubble-drop) interactions between them, we extend this study to their combined evolution through two arrangements: First, both are placed side-by-side in a row at the same height. We observe that the overall dynamic governs by the competition between the rotational motion induces due to the pairing between two co-rotating inner vorticity lobes and net vertical motion of two dipolar-vorticities (upward for bubble and downward for droplet) induces due to the gravity. In second arrangement where the vertically aligned bubble (below) and droplet (above) after collision exchange their blobs/lobes and subsequently start to move horizontally in opposite directions away from each other. This horizontal movement gets slower with increasing coupling strength. For the these arrangement, we consider the varying distance between the fixed-size bubble and droplet, and varying coupling strength which represents the nature of our medium. A series of 2D numerical simulations have been conducted to visualize these interactions. 
	\end{abstract}
	\maketitle
\section{Introduction}\label{Introduction}
%
 Rayleigh-Taylor (RT) instability occurs when an external force like gravity accelerates a heavy fluid into a lighter one \cite{rayleigh1900investigation,taylor1950instability,chandrhd}. Apart from RT instability, buoyancy-driven (BD) instability is another form of gravity-driven instability which have attracted considerable interest. The BD instability determines whether an object sinks or floats in the fluid  it is immersed in.  We have introduced the BD situation by including a localized low-density region (a bubble) and a localized high-density region (a droplet) than the fluid they are placed in. Under the influence of gravity, a bubble has a tendency to float upwards while a droplet sinks against the surrounding fluid. Bubbles and droplets have been studied separately, for example see \cite{shew2006viscoelastic,gaudron2015bubble,dollet2019bubble} for bubbles and see \cite{dwyer1989calculations,cristini2004theory,zhu2008three,leong2020droplet} for drops, and both together \cite{mokhtarzadeh1985dynamics,chen2008droplet,tabor2011homo} as well.  The knowledge of the interaction of gas bubbles and liquid droplets with themselves or with particles or with a colloid probe is desired for a wide range of applications, including the manufacturing of cosmetics and pharmaceuticals, oil recovery~\cite{zhao2017droplet}, printing and deinking, emulsion stability~\cite{xie2017surface}, foodstuffs such as ice cream and mousse~\cite{van2001aeration}, and in mineral flotation~\cite{liu2002fundamental,niewiadomski2007air}, among many other applications. Recently, detailed experiments~\cite{kong2019hydrodynamic} and numerical simulations~\cite{zhang2019vortex}  are conducted to investigate the interactions between a pair of bubbles rising side by side. Also, both homo- (bubble-bubble and drop-drop) and hetero- (bubble-drop) interactions between air bubbles and oil droplets measured by atomic force microscopy by~\cite{tabor2011homo}. 
 
    In part I of this paper, we separately explored the dynamics of a bubble and a droplet in the incompressible limit of the strongly coupled dusty plasmas (SCDPs).  We observed that a falling droplet process is equivalent to a rising bubble since both had symmetry in spatial distribution. We modeled the SCDP using an incompressible generalized hydrodynamic (i-GHD) fluid model~\cite{Kaw_Sen_1998, Kaw_2001}. This model treats SCDP as a viscoelastic (VE) fluid that characterizes the viscoelastic effects through two coupling parameters: shear viscosity $\eta$ and the Maxwell relaxation parameter $\tau_m$ \cite{frenkel_kinetic}. We observed that the rising  of a bubble and falling of a droplet get weaker with increasing coupling strength (ratio ${{\eta}/{\tau_m}}$).  In the present paper, to investigate the nature of interactions between a bubble and a droplet, we have both together  in the same fluid within the framework of i-GHD fluid model. We consider two arrangements: First, the droplet and bubble are placed side by side in a row at the same height; Second, the bubble and droplet are aligned in a column with the droplet placed above the bubble. To study these interactions, here, we consider the varying distance between the fixed-size bubble and droplet, and varying coupling strength which represents the nature of our medium.  

	This paper is organized in the following sections including the introduction as the first section.  The next Section~\ref{model_meth} describes the basic model equations and re-write them, under subsection `simulation methodology',  as a set of coupled equations for the implementation of our numerical scheme. In Section \ref{simulation_results}, to report simulation results, first we develop some qualitative understanding by using the model equations for the considered density profiles (bubble-droplet) through the schematic diagrams. For horizontally placed blobs, the simulations are performed for three initial spacing: widely spaced $(d{\gg}{2a_{c}})$, medium spaced $(d{>}{2a_{c}})$, and closely spaced $(d{\approx}{2a_{c}})$, ${a_{c}}$ is  core radius. Next, the droplet is placed above the bubble in a vertical column with fixed initial spacing. The role of coupling strength on bubble-droplet density dynamics has been depicted in the form of  transverse shear waves through respective vorticity contour plots. To develop a better physical insight into the dynamics of each phenomenon, the inviscid limit of HD fluids is also simulated for each case. Final Section~\ref{conclusions}, conclude the paper with a summary.

\section{The numerical model and simulation methodology} 
\label{model_meth}
The generalized hydrodynamics fluid model supports both the incompressible transverse and compressible longitudinal modes. To study the exclusive effect of transverse modes on the bubble and droplet interaction and to avoid the possible coupling with longitudinal mode, we consider the incompressible limit of dusty plasmas. In the incompressible limit, the Poisson equation is replaced by the quasi-neutrality condition and charge density fluctuations are ignored. The dust fluid flow under gravity acceleration $\vec{g}$ is characterized by a coupled set of continuity and momentum equations:
\begin{equation}\label{eq:continuity}
  \frac{\partial \rho_d}{\partial t} + \nabla \cdot
\left(\rho_d\vec{v}_d\right)=0{,}
  \end{equation}
 \begin{eqnarray}\label{eq:momentum0}
 &&\left[1+{\tau_m}\left(\frac{\partial}{\partial{t}}+{\vec{v}_d}\cdot \nabla\right)\right]\nonumber\\
 && \left[ {{\rho_d}\left(\frac{\partial{\vec{v}_d}}{\partial {t}}+{\vec{v}_d}{\cdot} \nabla{\vec{v}_d}\right)}+{\rho_d}\vec{g}+{\rho_c}\nabla \phi_{d} \right]\nonumber\\
 &&=\eta \nabla^2\vec{v}_d{,}
 \end{eqnarray}
respectively and the incompressible condition is given as
\begin{equation}\label{eq:incompressible}
 {\nabla}{\cdot}{\vec{v}_d}=0{.}
\end{equation}
  The derivation of these normalized equations has been discussed in detail in our earlier papers \cite{dharodi2014visco,dharodi2016sub} along with the procedure of its  numerical implementation and validation. Here,  ${\rho_d}= {n_d}{m_d}$ is the mass density of the dust fluid ($n_d$ is the number density of dust fluid, $m_d$ is the mass of the dust particle). The dust  charge density  ${\rho_c}= {n_d}{Z_d}$, $Z_d$ is the charge on each dust grain with no consideration  of charge fluctuation. The number density $n_d$ is normalized by the equilibrium value $n_{d0}$.  The dust charge potential $\phi_d$ is normalized by ${{K_B}{T_i}}/{e}$. The parameters $e$,  $T_i$ and $K_B$ are the electronic charge,  ion temperature and Boltzmann constant, respectively. The time, length, and dust fluid velocity  $\vec{v}_d$ are normalized by inverse of dust plasma frequency $\omega^{-1}_{pd}=\left({4\pi(Z_de)^{2}n_{d0}}/{m_{d0}}\right)^{-1/2}$, plasma Debye length $\lambda_{d}=\left({K_B T_i}/{4{\pi} {Z_d}{n_{d0}}{e^2}}\right)^{1/2}$, ${\lambda_d}{\omega_{pd}}$, respectively. In the hydrodynamic (HD) limit $i.e.$~$\tau_m{=}0$, this model represents a simple hydrodynamic fluid through the Navier-Stokes equation.
\subsection{Simulation methodology:} 
\label{num_methodology}
  For the numerical modeling the above generalized momentum Eq.~(\ref{eq:momentum0}) is transformed into a set of two coupled equations,
\begin{eqnarray}\label{eq:vort_incomp1}
{{\rho_d}\left(\frac{\partial{\vec{v}_d}}{\partial {t}}+{\vec{v}_d}{\cdot} \nabla{\vec{v}_d}\right)}+{\rho_d}\vec{g}+{\rho_c}\nabla \phi_{d}={\vec \psi}
\end{eqnarray}
\begin{equation}\label{eq:psi_incomp1}
\frac{\partial {\vec \psi}} {\partial t}+\vec{v}_d \cdot \nabla{\vec \psi}=
{\frac{\eta}{\tau_m}}{\nabla^2}{\vec{v}_d }-{\frac{\vec \psi}{\tau_m}}{.}
\end{equation}
We consider two-dimensional (2D) system lies in $xy$ plane, the $x$-coordinate is in the horizontal, $y$ in the vertical. Thus the above variables depend on  $x$ and $y$  $i.e$ ${\vec \psi}(x,y)$, ${\vec v_d}(x,y)$, and ${\rho_d}(x,y)$. The quantity ${\vec \psi}(x,y)$ is the strain produced in the elastic medium by the time-varying velocity fields. The density gradient and potential gradient are taken along the y-axis $i.e.$~${{\partial{\rho_d}}/{\partial y}}$, ${{\partial \phi_{d}}/{\partial y}}$, respectively.  The acceleration $\vec g$ is applied  opposite to the  fluid density gradient $i.e.$~$-g\hat{y}$.  We also assume no initial flow i.e. $\vec v_{d0}=0$ at $t=0$. With small perturbations; density,  scalar potential, and dust velocity can be written as 
\begin{equation}\label{eq:pert}
	{\rho_d}(x,y,t)={\rho_{d0}}(y,t=0)+{\rho_{d1}}(x,y,t) {,}
\end{equation}
\begin{equation}
	{\phi_d}(x,y,t)={\phi_{d0}}(y,t=0)+{{\phi_{d1}}}(x,y,t){,}
\end{equation}
\begin{equation}
	{\vec{v}_d}(x,y,t)=0+{\vec{v}_{d1}}(x,y,t) {,}
\end{equation}
respectively. Re-write the Eq.~(\ref{eq:vort_incomp1}) under the equilibrium condition, $ {\rho_{d0}}{g} =-{\rho_c}{{\partial\phi_{d0}} /{\partial y}}$ and using the perturbation density Eq.~(\ref{eq:pert}).
 \begin{eqnarray}\label{eq:vort_incomp2}
\frac{\partial{\vec{v}_d}}{\partial {t}}+{\vec{v}_d}{\cdot} \nabla{\vec{v}_d}+{\frac{\rho_{d1}}{\rho_d}}\vec{g}+{\frac{\rho_{c}}{\rho_d}}{\nabla{\phi_{d1}}}=\frac{\vec \psi}{\rho_d}{.}
 \end{eqnarray}
Taking the curl of Eq.~(\ref{eq:vort_incomp2}) and using the Boussinesq approximation  ($\rho_{d0}{\gg}\rho_{d1}$). The contribution from ${\nabla}{\phi_{d1}}$ becomes zero i.e. ${\nabla}{\times}{\nabla}{\phi_{d1}}=0$, we get  
\begin{equation}\label{eq:vort_incomp3} 
\frac{\partial{\xi_{z}}} {\partial t}+\left(\vec{v}_d \cdot \vec \nabla\right)
{\xi_{z}}={\frac{1}{\rho_{d0}}}{{\nabla}{\times}{\rho_{d1}}{\vec{g}}}+{\nabla}{\times}{\frac{\vec \psi}{\rho_d}}{.} 
\end{equation}
 $ {\xi_{z}}(x,y)={\vec \nabla}{\times}{\vec v_d}(x,y) $ is the vorticity which is normalised with dust plasma frequency. The final numerical model equations in term of variables x and y become 
\begin{equation}\label{eq:cont_incomp3}
 \frac{\partial \rho_d }{\partial t} +  \left(\vec{v}_d\cdot
\nabla\right)\rho_d= 0{,}
    \end{equation}
\begin{equation}\label{eq:psi_incomp3}
\frac{\partial {\vec \psi}} {\partial t}+\left(\vec{v}_d \cdot \vec
\nabla\right)
{\vec \psi}={\frac{\eta}{\tau_m}}{\nabla^2}{\vec{v}_d }-{\frac{\vec
\psi}{\tau_m}}{,}  
\end{equation}
\begin{equation}\label{eq:vort_incomp4} 
\frac{\partial{\xi}_z} {\partial t}+\left(\vec{v}_d \cdot \vec \nabla\right)
{{\xi}_z}=-{\frac{g}{\rho_{d0}}}{\frac{\partial{\rho_{d1}}} {\partial x}}
+{\frac{\partial}{\partial x}}\left({\frac{\psi_{y}}{\rho_d}}\right)
-{\frac{\partial}{\partial y}}\left({\frac{\psi_{x}}{\rho_d}}\right){.}   
\end{equation}

We use the {\small{LCPFCT}} package (Boris {\it et al.} \cite{boris_book}) to solve the set of  above coupled non-linear equations (\ref{eq:cont_incomp3}),~  (\ref{eq:psi_incomp3}) and (\ref{eq:vort_incomp3}) numerically. This package is based on the finite difference scheme associated with the flux-corrected agorithm. The velocity at each time step is updated by using the Poisson's equation ${\nabla^2}{\vec{v}_d}=-{\vec {\nabla}}{\times}{\vec \xi}$. This Poisson's equation has been solved by using the {\small{FISPACK}} \cite{swarztrauber1999fishpack}. Both of softwares are the package {\small{FORTRAN}} subprograms. Throughout simulation studies, boundary conditions are periodic in the horizontal direction (x-axis) while non-periodic  along the vertical (y-axis) direction where the effects of perturbed quantities die out before hitting the boundary of simulation box. In each case, the grid convergence study has been carried out first to make sure grid independence of the numerical results. The approximating the gravitational acceleration as $g=10$.

 In the hydrodynamic (HD) limit $i.e.$~$\tau_m{=}0$,  the Eq.~(\ref{eq:psi_incomp3}) has singularity. To avoid this singularity, we put ${\tau_m}$=0 in generalized momentum Eq.~(\ref{eq:momentum0}) and then take the curl, under the same above considered conditions and assumptions, one gets the HD vorticity equation
\begin{equation}\label{eq:vort_incomp_fluid} 
	\frac{\partial{\xi}_z} {\partial t}+\left(\vec{v}_d \cdot \vec \nabla\right)
	{{\xi}_z}=-{\frac{g}{\rho_{d0}}}{\frac{\partial{\rho_{d1}}} {\partial x}}+{\eta}{\nabla^2}{{\xi}_z}{.}   
\end{equation}
Thus, numerically, for pure HD cases ($\tau_m=0$) we solve the  set of  Eqs.~(\ref{eq:cont_incomp3}) and ~(\ref{eq:vort_incomp_fluid}) where dust fluid velocity at each time step is updated by using ${\nabla^2}{\vec{v}_d}=-{\vec {\nabla}}{\times}{\vec \xi}$.

\section{Simulation results}
\label{simulation_results}
\begin{figure}
	\includegraphics[width=1.0\textwidth]{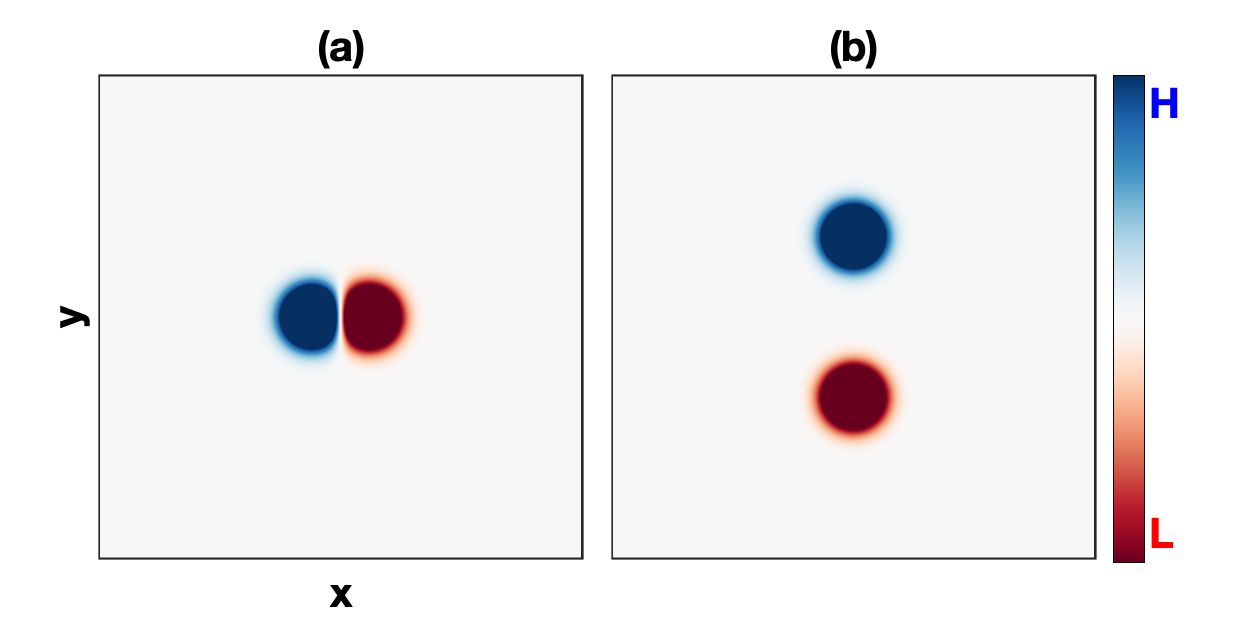} 
	\caption{The initial density profiles at time t=0 (a) the droplet (left) and the bubble (right) are placed at the same height, and (b) the droplet and the bubble are aligned vertically with the droplet placed above the bubble. In colorbar, letter {\bf {H}} is the acronym of the heavy density region and {\bf L} stands for lighter density region.} 
	\label{fig:figure1}	     
\end{figure}

To understand the bubble-droplet interactions, we consider two arrangements shown in  Fig.~\ref{fig:figure1}: Arrangement~({\small{A}}) in Fig.~\ref{fig:figure1}(a), the droplet (left) and bubble (right) are placed in a row at the same height; Arrangement~({\small{B}}) in Fig.~\ref{fig:figure1}(b), the bubble (bottom) and droplet (top) are aligned in a column.
\FloatBarrier
For both arrangements, the total density is $\rho_d=\rho_{d0}+\rho_{d1}$. ${\rho_{d0}}$ is the background density  and total density inhomogeneity is given by
\begin{equation}
\label{eq:net_prtbn_rel} 
 \rho_{d1}={\rho^{\prime}_{d1}+\rho^{\prime}_{d2}}{.}
\end{equation}
The Gaussian density inhomogeneity of the droplet centered at $({x_{c1}},{y_{c1}})$, with radius ${a_{c1}}$ is 
\begin{equation}
\label{eq:droplet_profile} 
\rho^{\prime}_{d1}={\rho^{\prime}_{01}}exp\left(\frac{-{\left(x-x_{c1}
		\right)^2-(y-y_{c1})^2}}{a^2_{c1}}\right){,}
\end{equation}
and the Gaussian density inhomogeneity of the bubble centered at $({x_{c2}},{y_{c2}})$, with radius ${a_{c2}}$ is 
\begin{equation}
\label{eq:bubble_profile} 
\rho^{\prime}_{d2}=-{\rho^{\prime}_{02}}exp\left(\frac{-{\left(x-x_{c2}
		\right)^2-(y-y_{c2})^2}}{a^2_{c2}}\right){.}
\end{equation}
Throughout the entire numerical simulations,  ${\rho_{d0}}=5$, ${\rho^{\prime}_{01}}$=${\rho^{\prime}_{02}}$=0.5, ${a_{c1}}={a_{c2}}=2.0$ are held constant. Thus, both the bubble and droplet have spatial symmetry and spaced by distance $d=\sqrt{({x_{c2}}-{x_{c1}})^2+({y_{c2}}-{y_{c1}})^2}$.

\begin{figure}
	\includegraphics[width=1.0\textwidth]{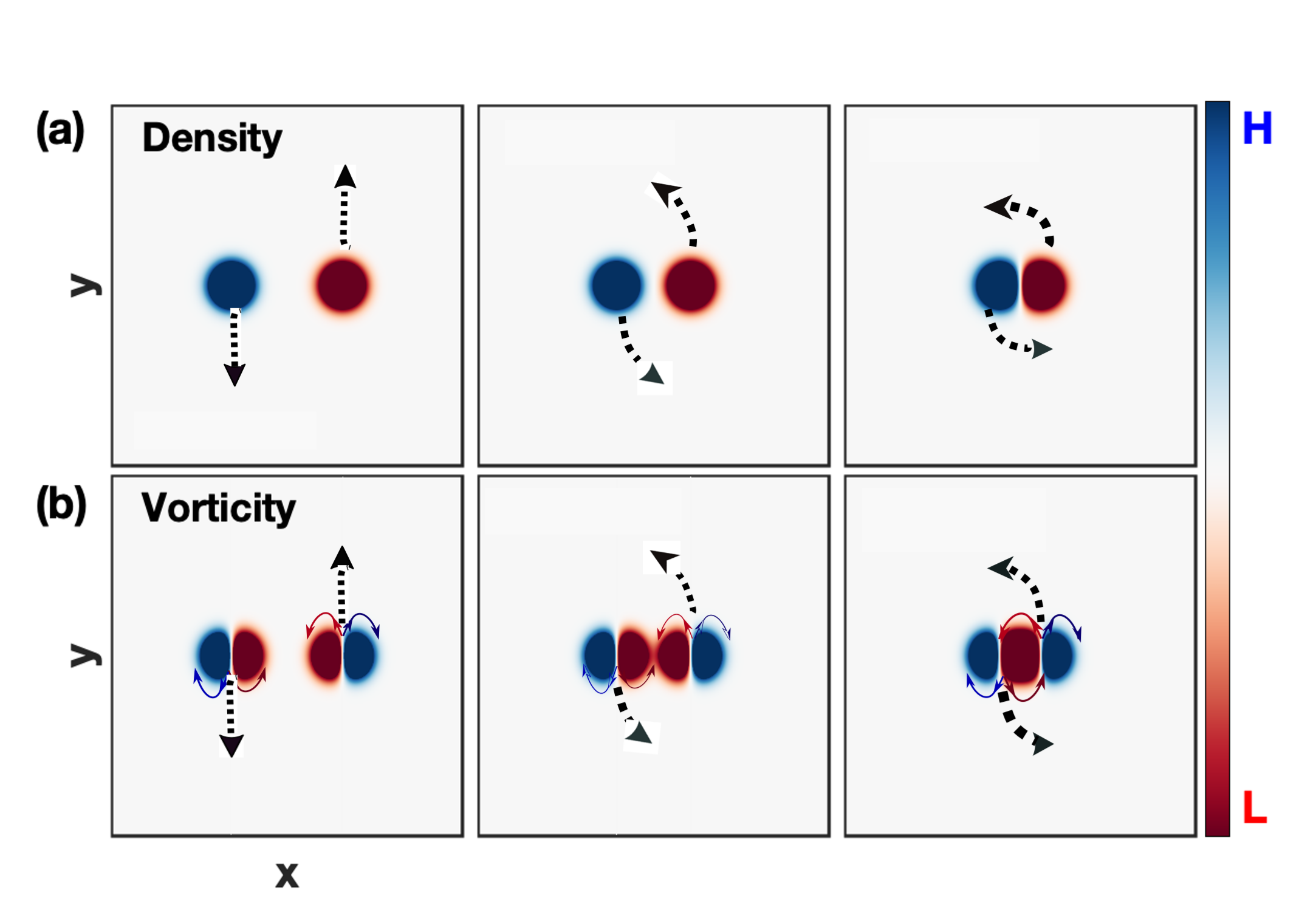}
	\caption{A schematic diagram of a droplet and a bubble placed in a row at the same height. Spacing between them decreases from left to right. The top row shows the density profile, the bottom row for the vorticity. The curved solid arrows represent the direction of rotation of the lobes in vorticity, and the net propagation is indicated by the dotted arrows.} 
	\label{fig:figure2}	  
\end{figure}

Before running codes, it would be helpful to develop some qualitative understanding by using the model equations for the considered density profile. For an inviscid HD  flow, the vorticity Eq.~(\ref{eq:vort_incomp4})/Eq.~(\ref{eq:vort_incomp_fluid}) using Eqs.~(\ref{eq:net_prtbn_rel}), ~(\ref{eq:droplet_profile}), and~(\ref{eq:bubble_profile}) becomes
\begin{equation}
\label{eq:fluid1} 
\frac{\partial{\xi_{z}}} {\partial t}+\left(\vec{v}_d \cdot \vec \nabla\right)
{\xi_{z}}={\frac{2g}{\rho_{d0}}}{\left(x-x_{c1}\right)}
{\rho^{\prime}_{d1}}-{\frac{2g}{\rho_{d0}}}{\left(x-x_{c2}\right)}{\rho^{\prime}_{d2}}{.}
\end{equation}
Here, the {\small{RHS}} represents the net vorticity  of combined bubble-droplet arrangement. Under the influence of gravity,  both terms depict the oppositely propagating dipolar vorticities each have a pair of counter-rotating (or unlike-sign) lobes. Thus, first term acts like a buoyant force on bubble causes vertical upward motion, while the second acts like a buoyant force on droplet causes vertical downward motion~\cite{horstmann2014wake}.  For the arrangement ({\small{A}}), using {\small{RHS}} of Eq.~(\ref{eq:fluid1}), Fig.~\ref{fig:figure2} represents the schematic images of vorticities  in the bottom row, corresponds to the bubble-droplet density profiles in the top row. The curved solid arrows over the lobes represent the direction of rotation of each lobe while the net propagation is indicated by vertical and curved dotted arrows.   
\FloatBarrier

In these schematic images, the spacing `$d$' between a droplet and a bubble decreases from left to right.  We observe that the overall dynamic governs by the competition between the rotational motion induces due to the pairing between two co-rotating inner vorticity lobes and net vertical motion of two dipolar-vorticities (upward for bubble and downward for droplet) induces due to the gravity. To understand it, first, let's discuss the case if the bubble and droplet are widely spaced (left-top panel), the counter-rotating vorticity lobes related to the bubble (left-bottom panel) causes vertical upward motion (indicated by vertical upward dotted arrow) while the vertical downward motion for the droplet (indicated by vertical downward dotted arrow). Due to large distance `$d$',  almost no interaction between the rising bubble and falling droplet, so their dynamics would be mainly governed by gravity $i.e.$  the same as individual cases. Whereas the decreasing separation distance (see rest of snapshots from left to right) would enhance the possibility of vorticity pairing through the co-rotating inner lobes. This pairing leads to the merging process \cite{von2000vortex,meunier2005physics,josserand2007merging} which results the  rotation (counterclockwise) of the entire structure about a common center of rotation.  Thus, this rotational effect is found proportional to the pairing of vorticity lobes. It is noted that, here we have the droplet on left side and bubble on right side means the co-rotating  (like-sign) inner lobes of net vorticity would rotate counterclockwise, otherwise clockwise for the reverse situation. 

For the arrangement ({\small{B}}), the droplet is aligned above the bubble in a vertical column as shown in Fig.~\ref{fig:figure1}(b). As discussed above, the generated dipolar vorticity related to the bubble causes the vertical upward motion while the vertical downward motion for the droplet. Thus, as time progresses, the two oppositely propagating dipolar vorticities collide with each other and then exchange their partners that result in two new dipolar structures that propagate away from each other in an orthogonal direction to the initial propagation. All the above discussed dynamics will be visualized through the numerical simulations in details in the subsequent sections. Thus far, these assessments are particularly true for an inviscid HD fluid where no  dissipation or source term exists.  In VE fluids, the vorticity Eq.~(\ref{eq:vort_incomp3}) or Eq.~(\ref{eq:vort_incomp4}) besides the gravity term includes an additional term ${\nabla}{\times}{({\vec \psi}/{\rho_d})}$ in the {\small{RHS}}. This term incorporates the TS waves emerging from the rotating lobes into the medium. The speed of TS waves ($\sqrt{{\eta}/{{\tau_m}}}$) is proportional to the coupling strength of the medium.   For a fixed $d$, the effect of VE nature on the bubble-droplet interactions has been introduced through the varying coupling strength of the medium.
\subsection{ Aligned horizontally}
\label{align_hrzntl}
We have considered a system of length $lx=ly=24{\pi}$ units with $512{\times}512$ grid points in both the x and y directions. The system along the x-axis and y-axis is from $-12{\pi}$ to  $12{\pi}$ units.  The droplet and bubble are placed at the same height $({y_{c1}}, {y_{c2}})=(0, 0)$. It is worth noting at this point that the interaction between a bubble and a droplet depends, apart from the nature of a medium, on their sizes $i.e.$~${a_{c1}}$/${a_{c2}}$ (both have equal and fixed,  ${a_{c}}={a_{c1}}={a_{c2}}=2.0$), on their  shapes (both are symmetric), and initial spacing between them $d={x_{c2}}-{x_{c1}}$. Here, the simulations are performed for three spacing: widely spaced $(d{\gg}{2a_{c}})$, medium spaced $(d{>}{2a_{c}})$, and closely spaced $(d{\approx}{2a_{c}})$. The separation distances are $d=12$ with $({x_{c1}},{x_{c2}})=(6.0,-6.0)$, $d=8$ with $({x_{c1}},{x_{c2}})=(4.0,-4.0)$, and $d=4.4$ with $({x_{c1}},{x_{c2}})=(2.2,-2.2)$, for cases (i), (ii), and (iii) respectively.  In each case, the coupling strength has been introduced as the mild-strong ($\eta$=2.5, $\tau_m$=20), medium-strong ($\eta$=2.5, $\tau_m$=10) and strong or strongest ($\eta$=2.5, $\tau_m$=5), and pure viscosity ($\eta$=2.5, $\tau_m$=0).
\subsubsection*{Case (i): Widely spaced $(d{\gg}{2a_{c}})$}
\begin{figure}
	\includegraphics[width=1.0\textwidth]{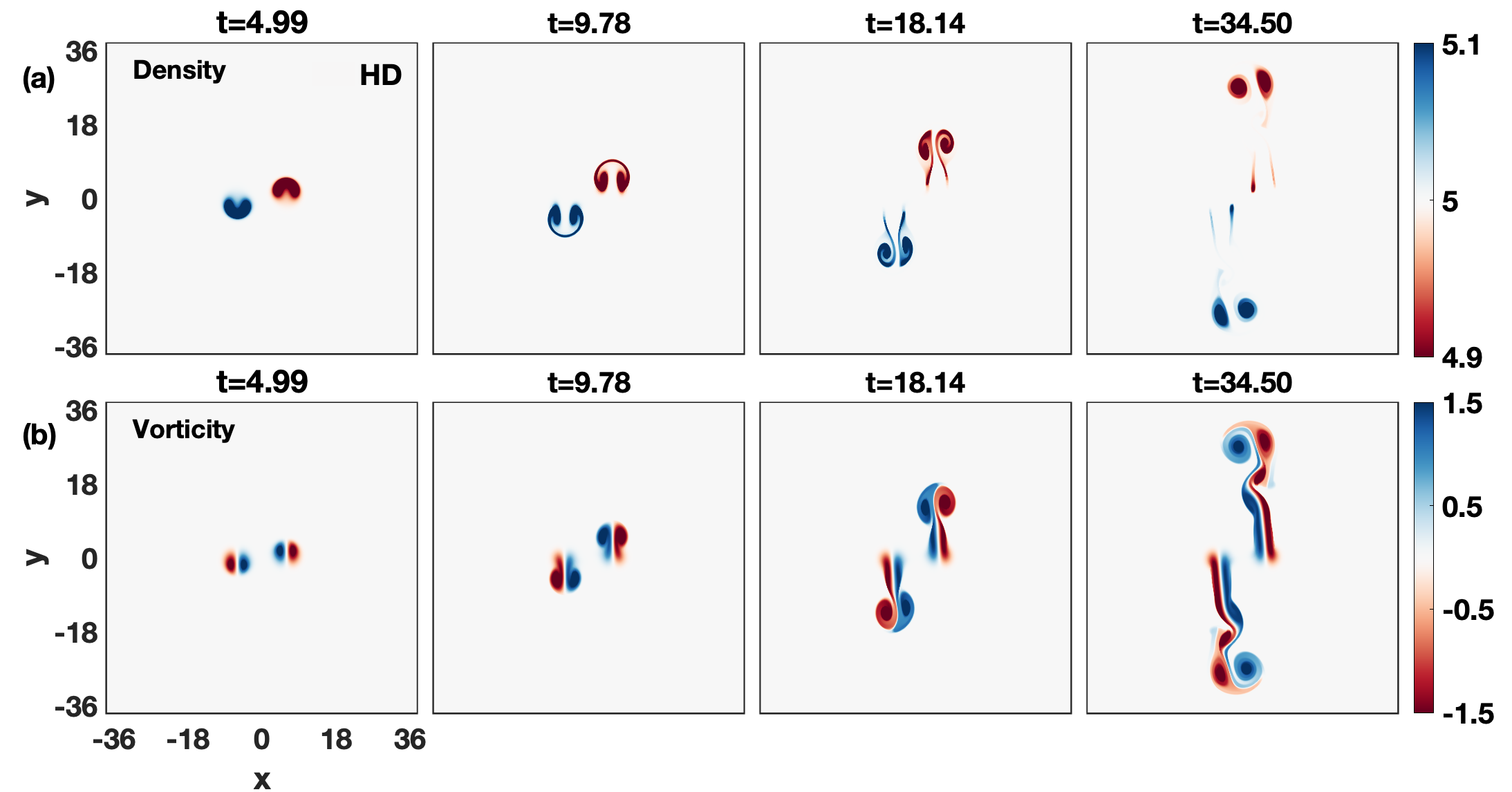}
	\caption{Time evolution of bubble-droplet density (top row) and vorticity (bottom row) for an inviscid hydrodynamic fluid. Both are separated by distance $d=12$ units  $(d>>2a_c)$. The dynamics of the rising bubble and falling droplet are almost independent and evolve the same as the individual ones.}
	\label{fig:figure3}	      
\end{figure}

Initially, the droplet (${x_{c1}}=-6.0$) and bubble (${x_{c2}}=6.0$) are quite well spaced $(d=12>>2a_c, {a_{c}}=2)$ without any overlap between the inner lobes of vorticities. First, let's understand this bubble-droplet dynamics for an inviscid HD fluid. Here, the absence of any dissipation term  makes this combined evolution like an individual one. In part I of this paper, we have studied a rising bubble and a falling droplet separately. We observed the falling droplet process is equivalent to the rising bubble as they initially have axisymmetry in density profiles. Owing to gravity, the initially circular density blobs change into crescent shapes. Further, as time progresses, both the bubble and droplet blobs break up into two distinct density blobs. The two blobs of the bubble/droplet propagate upward/downward as a single entity leaving behind the wake-like structure in background fluid. Such observations are clearly evident from the time evolution of the combined density profile for both in Fig.~\ref{fig:figure3}(a). The reason for the rising/falling bubble/droplet can be visualized from vorticity evolution in Fig~\ref{fig:figure3}(b). As the simulation begins, the buoyant forces are induced in the form of oppositely propagating two dipolar vorticities each has two counter-rotating lobes. The resultant is the vertical upward motion for a bubble while the droplet moves vertically downward, as discussed in detail above.  
\FloatBarrier

Next, it would be interesting to see this bubble-droplet dynamic in {\small{SCDP}s} which are depicted as {\small{VE}} fluids. The {\small{SCDP}s}, in addition to the forward motion, favor the radial emission of {\small{TS}} waves into the ambient fluid from each rotating vorticity lobes. These waves have the same symmetry of a lobe, and their speed is proportional to the coupling strength (${{\eta}/{\tau_m}}$) of the medium. In other words, a medium with stronger coupling strength would support the faster {\small{TS}} waves. The faster wave travels a greater distance in the same amount of time and that results in a faster spreading of lobes. Thus, for the current horizontal arrangement, a medium with strong coupling strength shows the higher probability of vorticity interaction or pairing between two like-signed inner lobes at an earlier time which in turn would enhance the rotational effect in the medium. In addition, the emerging TS waves from both the outermost lobes also help in enhancing this interaction by pushing the inner lobes towards each other. All this suggests that the competition between the rotational strength of two inner like-sign vorticity lobes due to the lateral interaction and forward vertical motion of two dipolar vorticities (unlike-sign lobes) due to the gravity governs the net dynamics. In the subsequent section, only the coupling strength is changing through elastic term $\tau_m$ for the fixed viscosity $\eta=2.5$. 

\begin{figure}
	\includegraphics[width=1.0\textwidth]{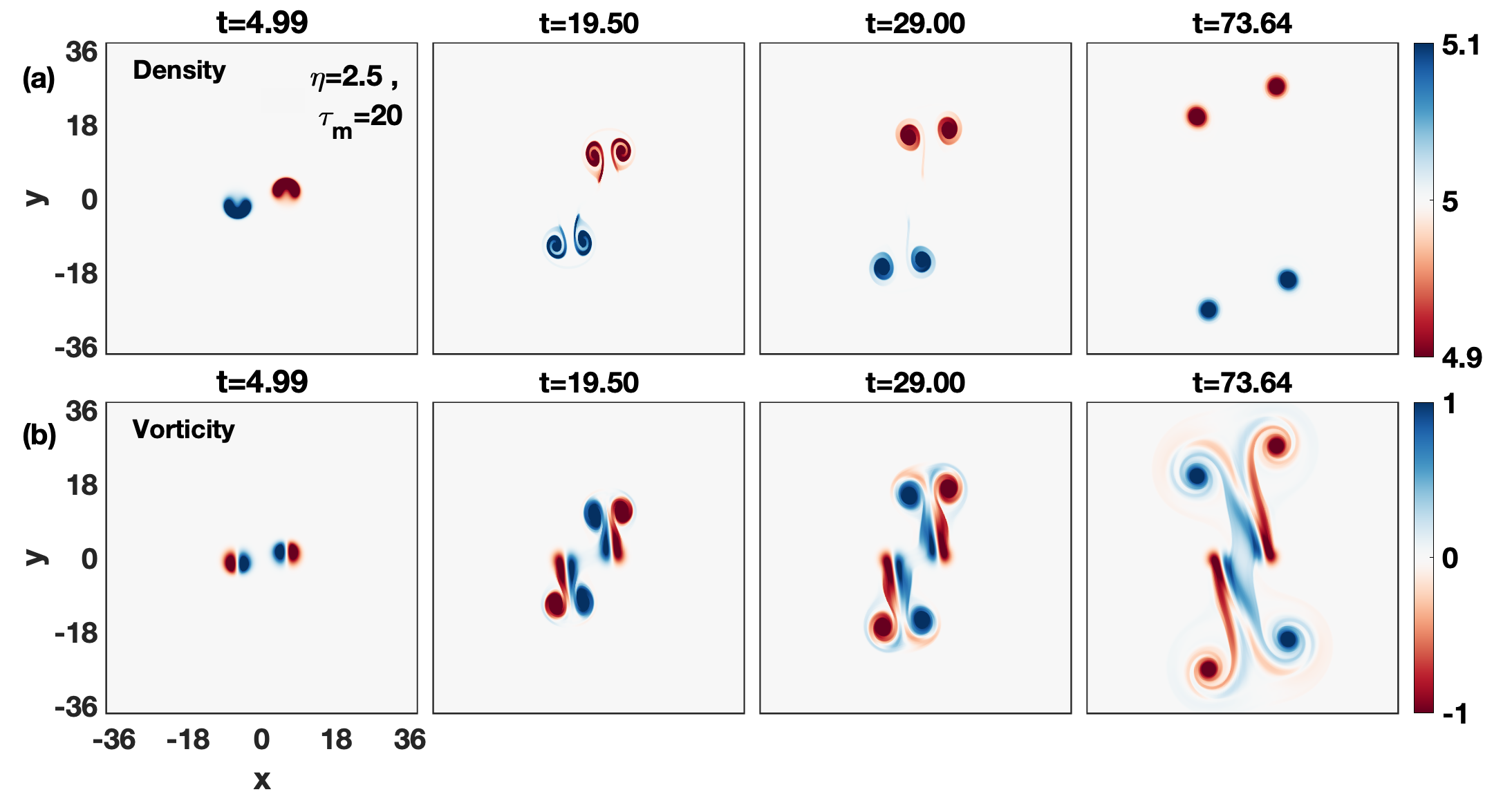}
	\caption{Time evolution of bubble-droplet density (top row) and vorticity (bottom row) for a viscoelastic fluid with $\eta=2.5$ and $\tau_m$=20. In comparison to the {\small{HD}} fluid (Fig.~\ref{fig:figure3}), there is a decrease in the rising/falling rate of the bubble/droplet and an increase in horizontal separation between the lobes.}
	\label{fig:figure4}	      
\end{figure}%
Let's start with mild-strong case $i.e.$~$\eta=2.5, \tau_m$=20 as shown in Fig.~\ref{fig:figure4}.  Similar to the Fig~\ref{fig:figure3}, here also the forward vertical motion due to the gravity dominates over the lateral interaction. But, in addition, it is evident from  Fig.~\ref{fig:figure4}(b) that there is emission of TS waves surrounding the vorticity lobes and no such waves exist in Fig.~\ref{fig:figure3}(b). These TS waves result in a mutual pushing between unlike-signed lobes which in turn enhances the separation between them with time. Besides lobes separation, the TS wave reduced the strength of dipoles thereby reducing their propagation. The relative observations of Fig.~\ref{fig:figure4}(a) and Fig~\ref{fig:figure3}(a) clearly reflect the aforementioned fact. Next Fig.~\ref{fig:figure5} shows the evolution of density (top row) and vorticity (bottom row) for the medium-strong coupling strength $\eta$=2.5, $\tau_m$=10. Also in Fig.~\ref{fig:figure5}(b), the forward vertical motion dominates over the lateral interaction. But due to the higher coupling strength, the reduction in forwarding motion and enlargement in horizontal separation between the unlike-sign dipoles are higher in comparison to the earlier case (Fig.~\ref{fig:figure4}; $\eta=2.5, \tau_m$=20).
\begin{figure}
	\includegraphics[width=1.0\textwidth]{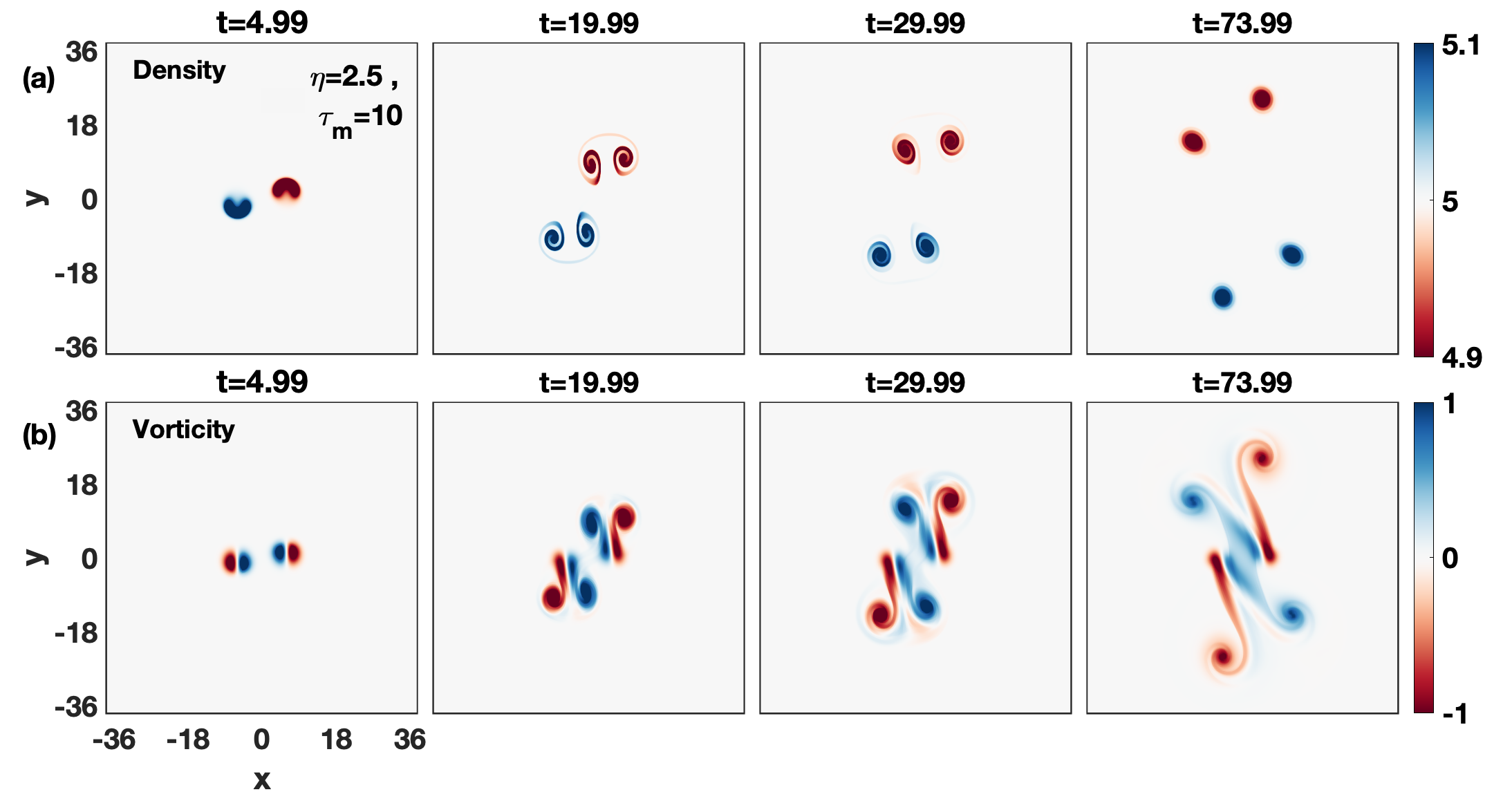}
	\caption{Time evolution of bubble-droplet density (top row) and vorticity (bottom row) for a viscoelatic fluid with $\eta=2.5$ and $\tau_m$=10. Both the density blobs  are separated by distance d=12 units  $(d{>>}2a_c, a_c=2)$.}
	\label{fig:figure5}	      
\end{figure}%
A viscoelastic fluid with a strong coupling strength ($\eta=2.5, \tau_m$=5.0) than earlier cases is shown in Fig.~\ref{fig:figure6}. It would support the even faster TS waves. So, this in addition to slow down the forward motion of lobes will enhance the vorticity exchange between the inner lobes at the earlier time. From Fig.~\ref{fig:figure6}(b), it is clearly evident that the rotation between inner lobes due to the merging process dominates over the forward vertical motion. As time passes, the combined effect of motion under gravity and the pairing of inner lobes causes the continuous stretching and spreading of inner lobes (see the second and third vorticity snapshots). The inner expanding lobes rotate around each other for a while and merge into an elliptic vortex at later times (forth vorticity snapshots). Here, because of emission of TS waves the rotating elliptic vortex grows in time and the merging occurs without the need of a third vortex \cite{kevlahan1997vorticity}. 
\begin{figure}
	\includegraphics[width=1.0\textwidth]{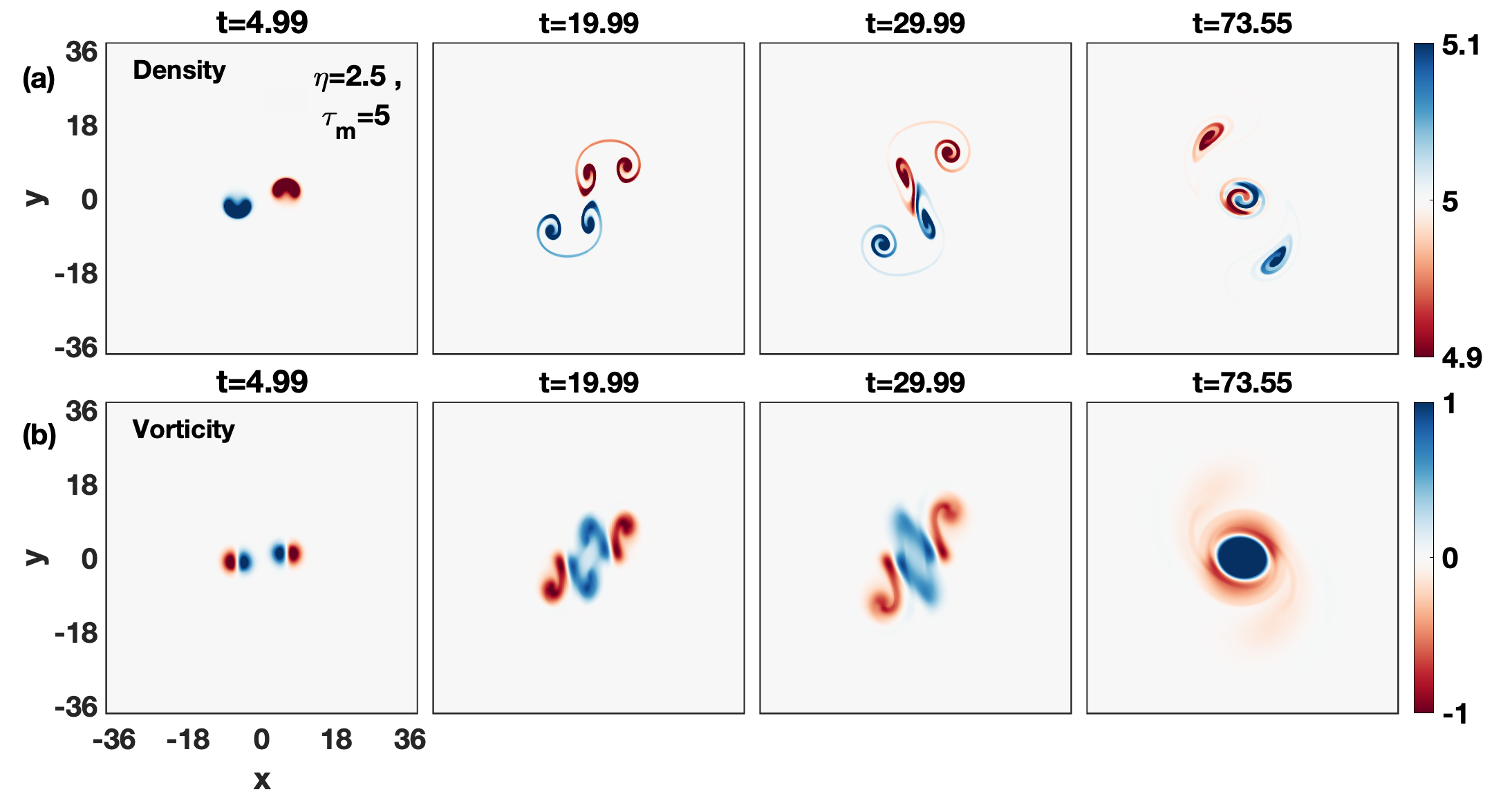}
	\caption{Time evolution of bubble-droplet density (top row) and vorticity (bottom row) for a viscoelatic fluid with $\eta=2.5$ and $\tau_m$=5.  Both the density blobs  are separated by distance d=12 units  $(d{>>}2a_c, a_c=2)$. The rotation between inner lobes due to the merging process dominates over the forward vertical motion. }
	\label{fig:figure6}	      
\end{figure}%
In Fig.~\ref{fig:figure6}(a), the initially circular density blobs change into crescent shapes. Due to the reduction in the vertical forward motion and enhancement in the transverse direction, these crescent shapes get wider in the horizontal direction for a while and then transform into curved structures with dumbbell-shaped edges of blobs.  (see the second density snapshot) unlike inviscid HD fluid (Fig.~\ref{fig:figure3}(a)). Later, the merging/pairing between two inner blobs of different densities form a center vortex and separates it from two outer blobs results in a tripolar structure (see the third density snapshot). This tripolar structure has two outer blobs (one from a bubble and one from a droplet) revolving around the center vortex like satellites. Thus, it has a center vortex with two rolling distinct density arms and two outer blobs spiral into one arm (see the forth density snapshot). This observation favors our recent study of spiral waves in density hetrogeneous viscoelastic fluids that the number of spiral arms is proportional to the number of different densities coexist \cite{dharodi2020rotating}. This tripole is a symmetric tripole as it does not move merely rotate \cite{van1991formation}.

\begin{figure}
	\includegraphics[width=1.0\textwidth]{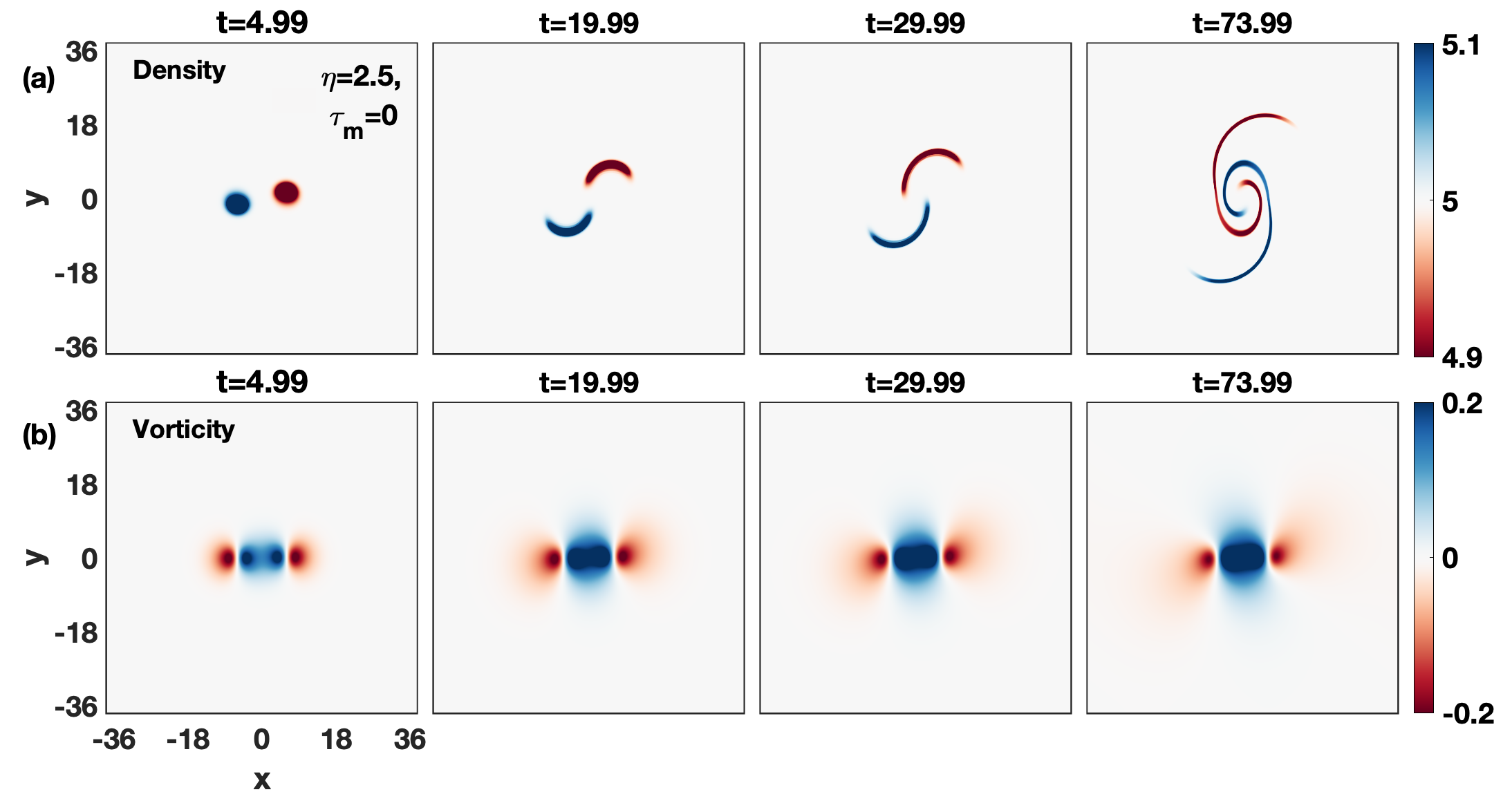}
			\caption{Time evolution of bubble-droplet density (top row) and vorticity (bottom row) for pure viscous HD fluid ($\eta$=2.5; $\tau_m$=0). Here, due to the viscous damping force,  in the persence of gravity the density crescent structures is transformed into spirals with time.}	
	\label{fig:figure7}	      
\end{figure}

In order to see the pure viscous effect of viscosity on bubble-droplet dynamics, we have also simulated a pure viscous HD fluid (see the density evolution in Fig.~\ref{fig:figure7} for $\eta$=0.1; $\tau_m$=0) by using Eqs.~(\ref{eq:cont_incomp3}) and ~(\ref{eq:vort_incomp_fluid}). In Fig.~\ref{fig:figure7}(b) the viscous effect is so strong over the gravity effect that there is almost no forward vertical motion. The spreading sizes of vorticity lobes increase the attraction between the like-sign inner lobes with time by diffusion~\cite{huang2005physical} and this, in turn, induces a regularly rotating flow at the center.
This reduction in the vertical motion and strong rotating flow at the center transformed the crescent density blobs into two persistently rotating spirals with time as shown in Fig.~\ref{fig:figure7}(a). Again, this observation confirms that the number of spiral arms is proportional to the number of different densities coexist~\cite{dharodi2020rotating}. 

Thus, the bubble-droplet dynamics are significantly affected by the presence of $\tau_m$ which controls the viscous spreading of vorticity lobes through the existence of transverse mode in SCDPs.

\subsubsection*{Case (ii): Medium spaced $(d{>}{2a_{c}})$}
\begin{figure}
	\includegraphics[width=1.0\textwidth]{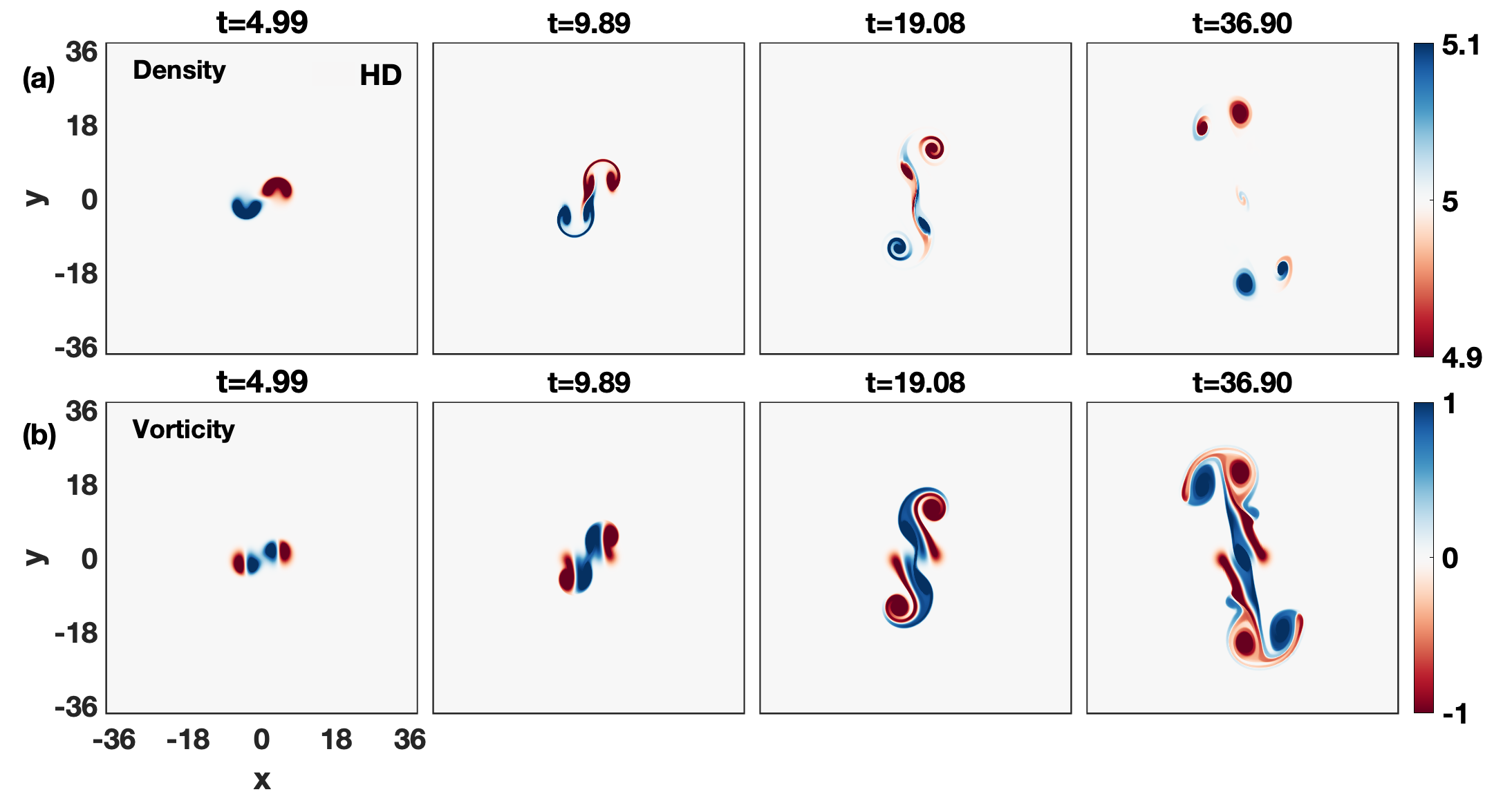}
	\caption{Time evolution of bubble-droplet density (top row) and vorticity (bottom row) for an inviscid hydrodynamic fluid. Initially, the droplet and bubble have a small overlapping between the inner lobes. Both have some interactions in the beginning but later start to evolve independently as the separate one with some diversion from the vertical direction in comparison to the widely spaced HD case shown in Fig.~\ref{fig:figure3}.}
	\label{fig:figure8}	      
\end{figure}

At the start, the droplet (${x_{c1}}=-4.0$) and bubble (${x_{c2}}=4.0$) are medium spaced $(d=8)$, two inner lobes of vorticities just touch each other. Here, spacing is less than the earlier widely spaced case ($d=12$). Thus, there should be a higher probability of interaction between a bubble and a droplet. Fig.~\ref{fig:figure8} displays the time evolution of the density (top row) and vorticity (bottom row) profiles for an inviscid hydrodynamic fluid.
\FloatBarrier
In the beginning, Fig.~\ref{fig:figure8} shows a small side-by-side overlapping between the inner lobes due to some closeness between them. Later, the forward motion (induces due to the gravity) of a bubble and a droplet dominates over this transverse overlapping and results to evolve them independently as the separate one with some diversion to the vertical direction in comparison to the earlier widely spaced HD case (Fig.~\ref{fig:figure3}).

\begin{figure}
	\includegraphics[width=1.0\textwidth]{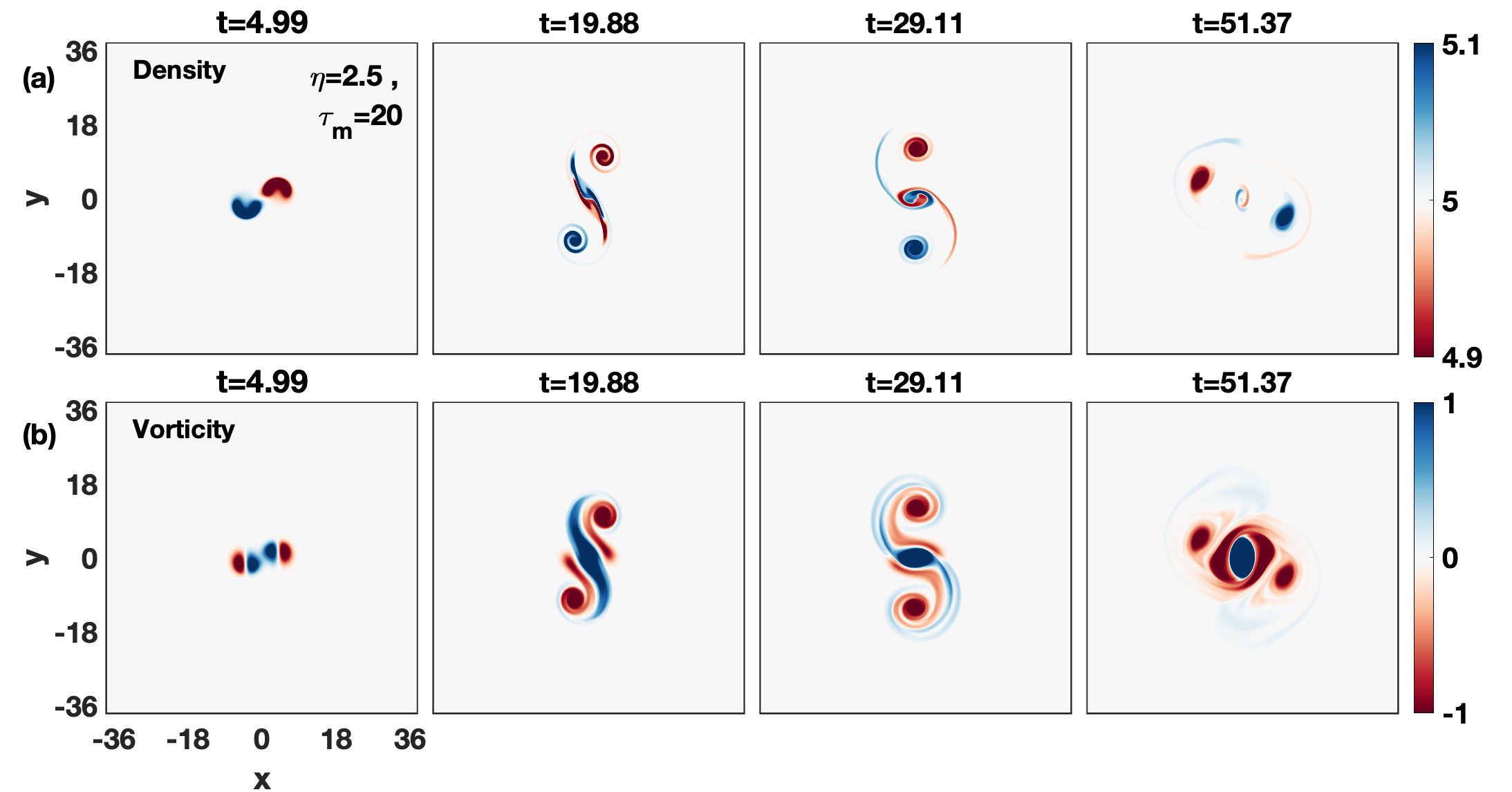}
		\caption{Time evolution of bubble-droplet density (top row) and vorticity (bottom row) for a viscoelastic fluid with $\eta=2.5$ and $\tau_m$=20.  Both are  separated by distance d=8.0 units  $(d{>}2a_c, a_c=2)$}
	\label{fig:figure9}	      
\end{figure}

 In order to visualize the effect of coupling strength, let's again start with the mild-strong coupling case; $\eta=2.5, \tau_m$=20 in Fig.~\ref{fig:figure9}. 
In Fig.~\ref{fig:figure9}(a,b) the merging process dominates over the forward vertical motion and this results the formation of tripolar structure. While for the earlier widely spaced case ($d=12$) the tripolar structure formed only for a strong coupling strength  ($\eta$=2.5; $\tau_m$=5). Thus, the decreasing distance increases the attraction between the bubble and droplet. 

\begin{figure}
	\includegraphics[width=1.0\textwidth]{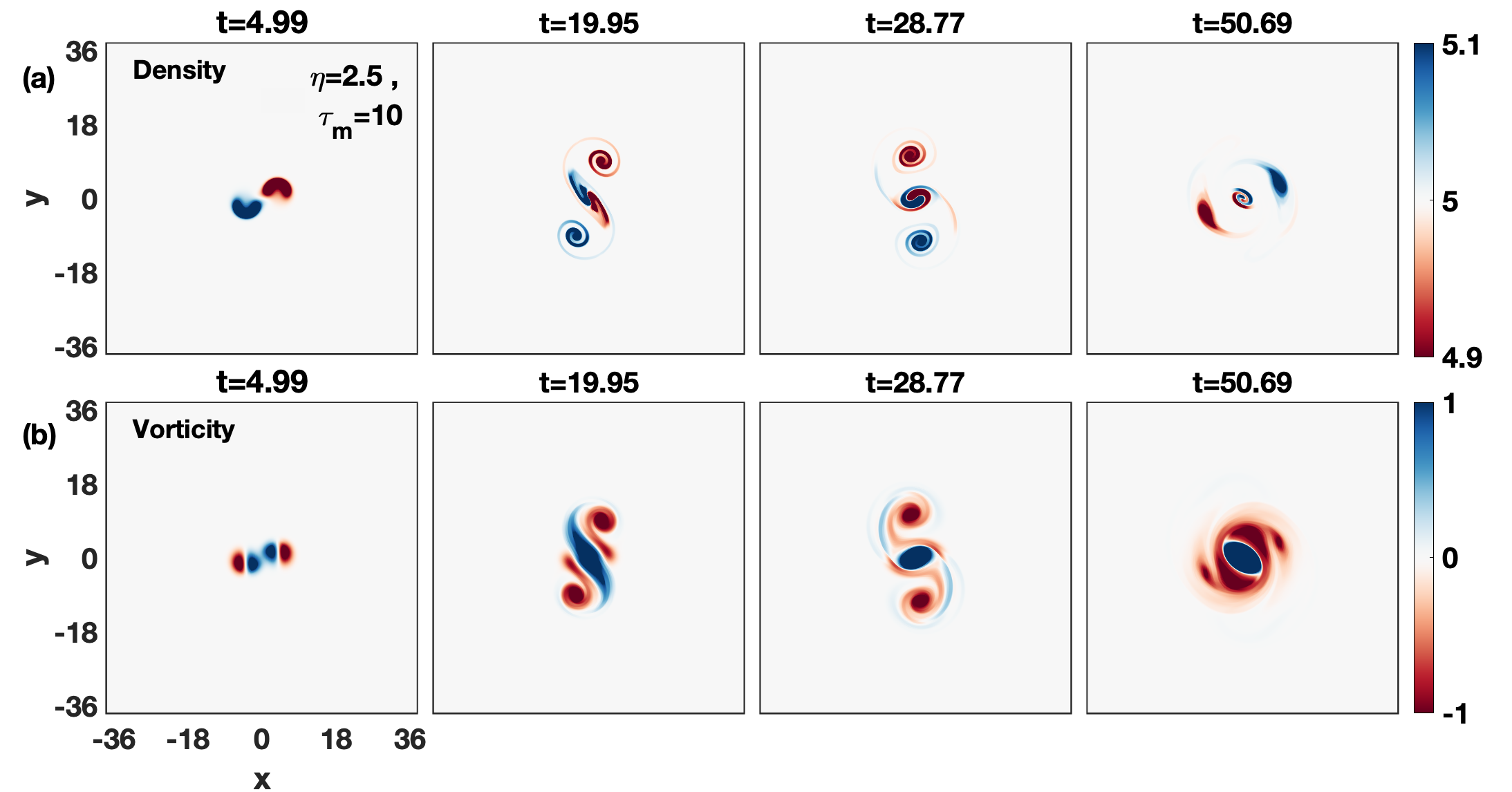}
		\caption{Time evolution of bubble-droplet density (top row) and vorticity (bottom row) for a viscoelastic fluid with $\eta=2.5$ and $\tau_m$=10.}
	\label{fig:figure10}	      
\end{figure}

\begin{figure}
	\includegraphics[width=1.0\textwidth]{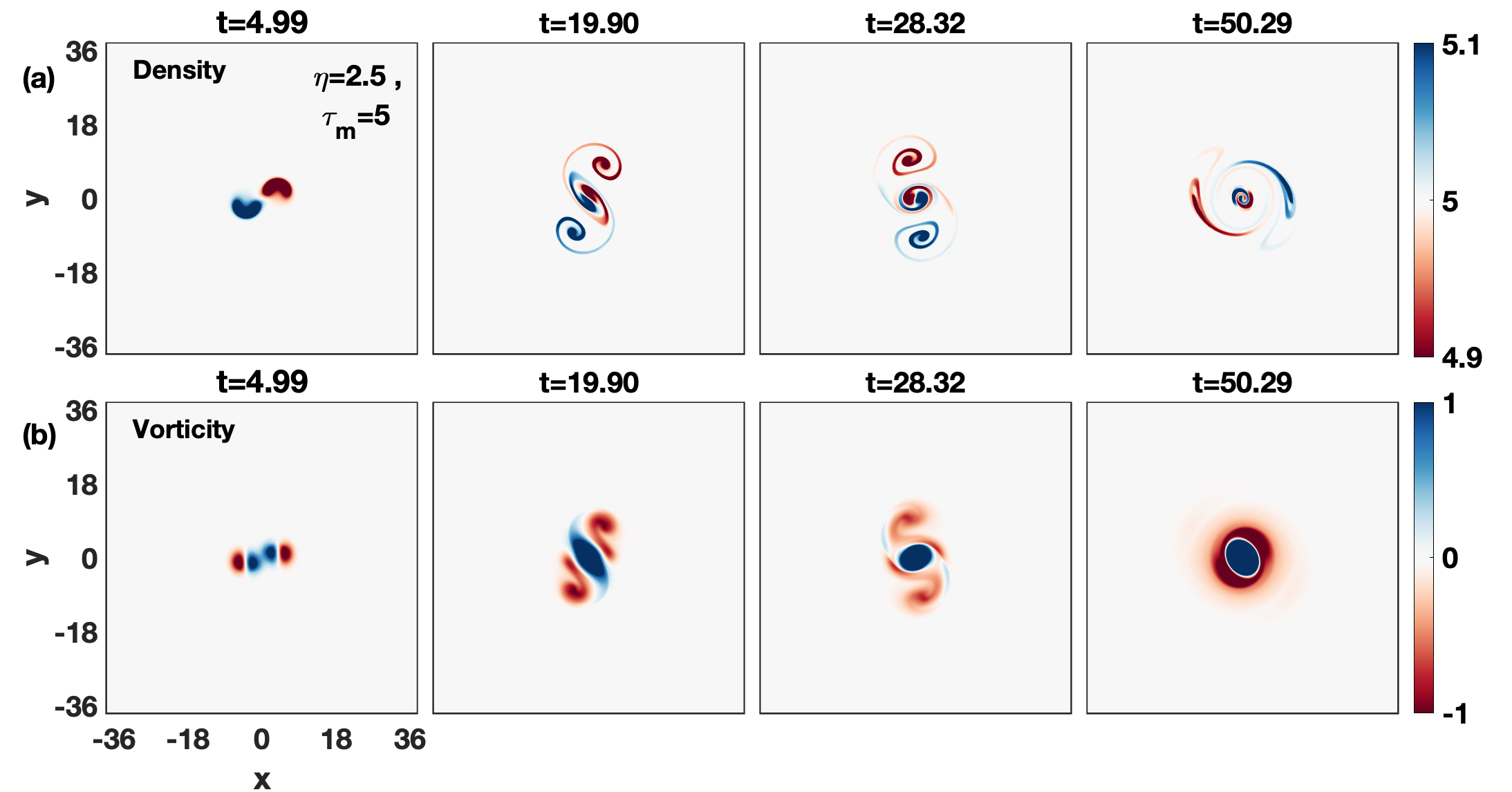}
		\caption{Time evolution of bubble-droplet density (top row) and vorticity (bottom row) for a viscoelastic fluid with $\eta=2.5$ and $\tau_m$=5. Both are  separated by distance d=8.0 units  $(d{>}2a_c, a_c=2)$}
	\label{fig:figure11}	      
\end{figure}

\begin{figure}
	\includegraphics[width=1.0\textwidth]{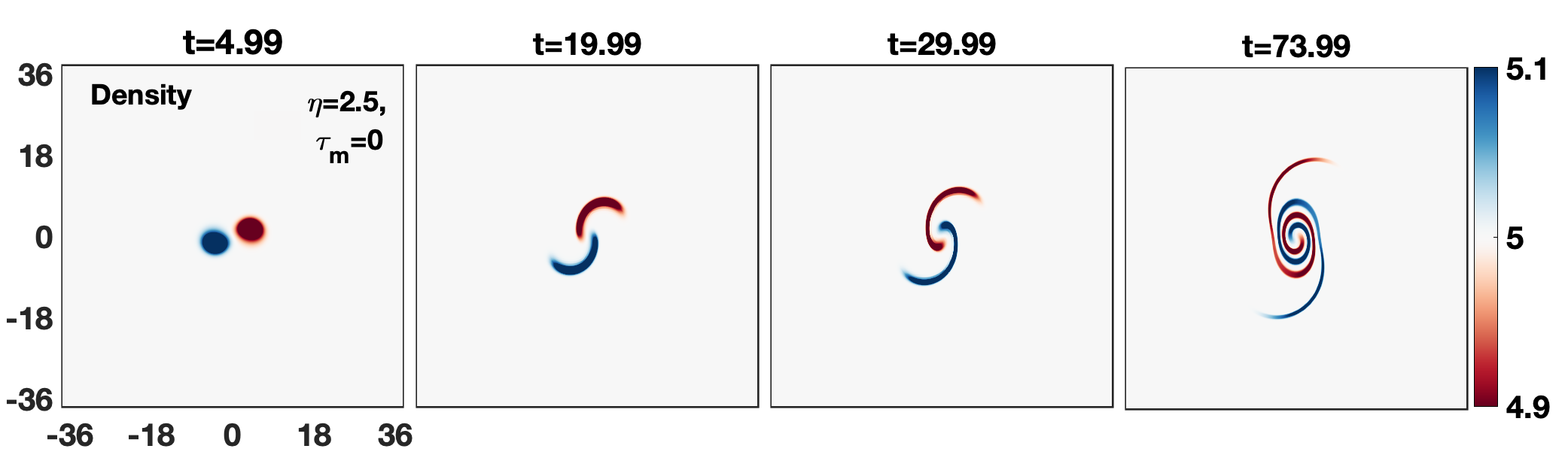}
	\caption{Time evolution of bubble-droplet density (top row) and vorticity (bottom row) for pure viscous HD fluid ($\eta$=2.5; $\tau_m$=0). This medium spaced blobs ($d{>}2a_c$) shows the higher rotaion rate of spirals than widely spaced case $d{>>}2a_c$ (Fig.~\ref{fig:figure7}).}	
	\label{fig:figure12}	      
\end{figure}

Next, for the medium-strong case $\eta=2.5; \tau_m$=10 in Fig.~\ref{fig:figure10}(a,b), the formation of a tripolar structure takes place prior in comparison to the earlier case ($\eta=2.5, \tau_m$=20). This tripolar structure has a smaller size and high rotation speed. The smaller size means less area of convection of the fluid across the medium. The rotating outer density blobs are also no longer circular.
For the case of strong coupling strength $\eta=2.5; \tau_m$=5.0, Fig.~\ref{fig:figure11}, the reduction in vertical motion is higher and the net structure rapidly merges into a tripolar. As time passes, the outer vorticity lobes arms are spread out around the inner rotating elliptical vortex and smooth out by the emerging TS shear waves. Thus, more axial symmetric structure formation takes place.
A comparative analysis of all the above cases shows that the confinement of the tripolar structures is proportional to the increasing coupling strength. Figure~\ref{fig:figure12} shows the evolution of bubble-droplet density of a pure viscous fluid with $\eta$=2.5; $\tau_m$=0. Under the influence of gravity, in the absence of the TS wave, the crescent structures transformed into two regularly rotating spirals. These spirals are elongated vertically due to the gravitational force. 
It also observed from the comparison of Fig.~\ref{fig:figure12} and Fig.~\ref{fig:figure7}(a) that the medium spaced blobs ($d{>}2a_c$) shows the higher rotation rate of spirals than widely spaced ($d{>>}2a_c$).

\subsubsection*{Case (iii): Closely spaced $(d{\approx}{2a_{c}})$}
To begin with, the droplet (${x_{c1}}=-2.2$) and bubble (${x_{c2}}=2.2$) are placed close enough $(d=4.4\approx2a_c)$ that both the like-sign inner lobes of vorticities almost overlap with each other this yields a tripolar vorticity structure. As the simulation begins, under the influence of gravitational force, the center common vortex starts to rotate counter-clockwise and this results a complete rotational flow about a common center even for an inviscid hydrodynamic fluid as shown in Fig.~\ref{fig:figure16}.  
\begin{figure}
	\includegraphics[width=1.0\textwidth]{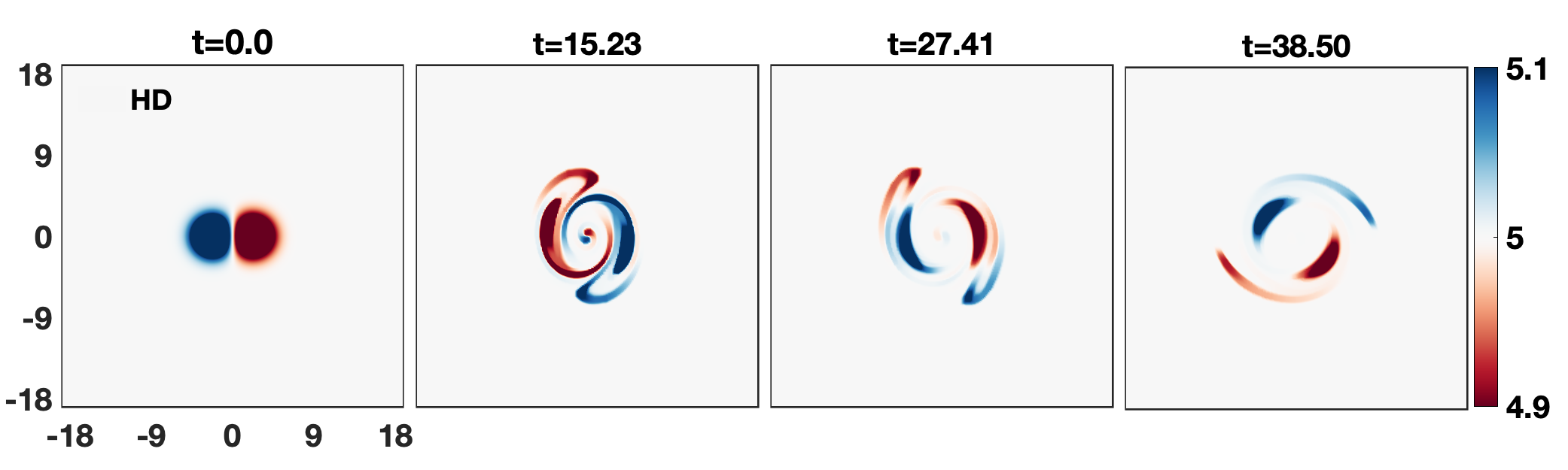}
	\caption{Time evolution of bubble-droplet density for an inviscid hydrodynamic  fluid.  Both are  separated by distance d=4.4 units  $(d{\approx}2a_c, a_c=2)$.}
	\label{fig:figure13}	      
\end{figure}%
\FloatBarrier
This rotating flow transforms the crescent-shaped density blobs into thin intertwining spirals as shown in Fig.~\ref{fig:figure13}. Up to $t{\approx}15$, there is no significant difference in density profiles for simple fluid and GHD/viscoelastic cases.  This common evolution features can be clearly seen up to second snapshot of Figs.~\ref{fig:figure14}(a),~\ref{fig:figure14}(b), and ~\ref{fig:figure14}(c) for varying  relaxation parameter $\tau_m$=20, 10, and 5, respectively; for the fixed viscosity $\eta$=2.5.
\begin{figure}
	\includegraphics[width=1.0\textwidth]{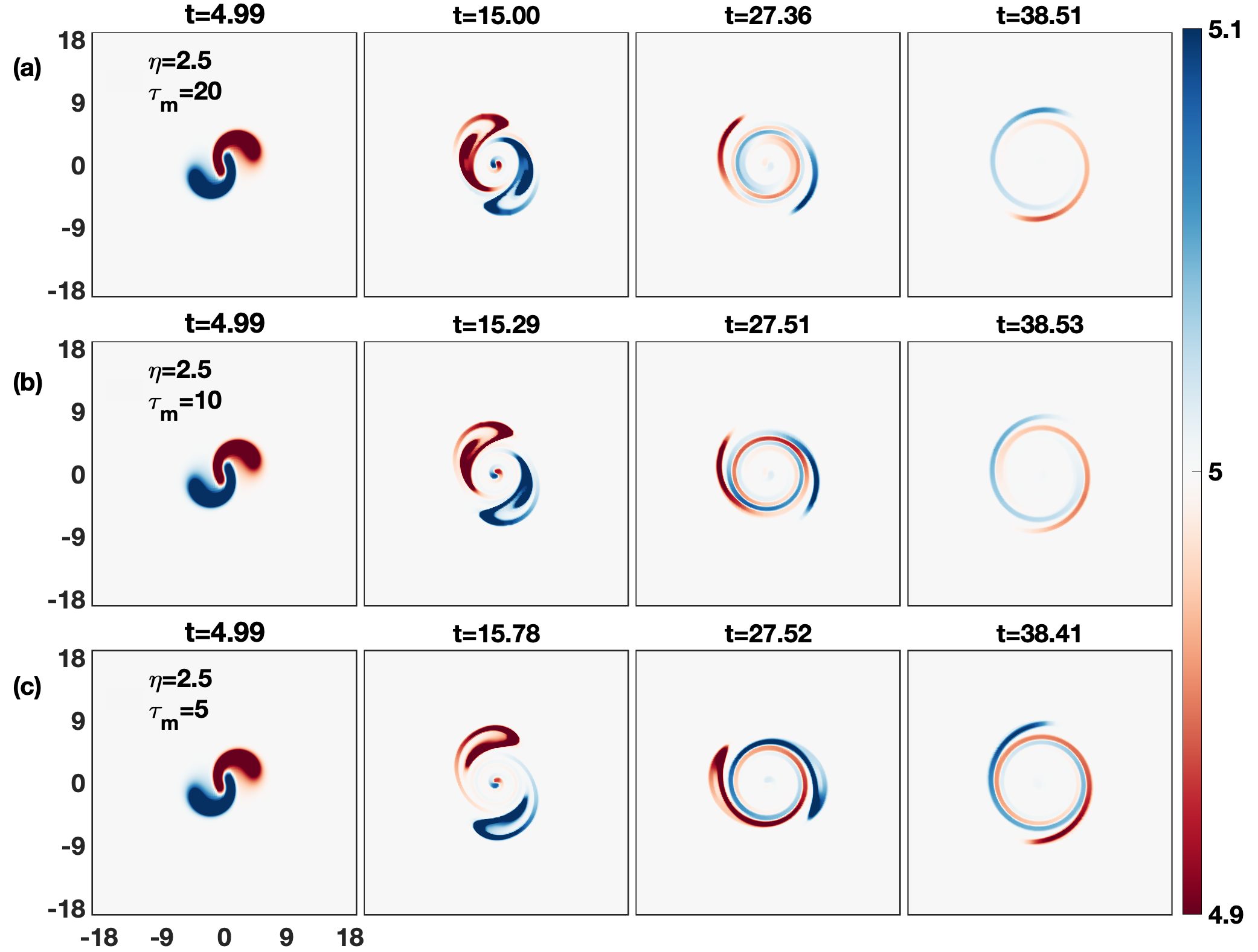}
	\caption{Time evolution of bubble-droplet density  in viscoelastic fluids for fixed viscous term $\eta$=2.5; and varying coupling strength due to changing relaxation parameter $\tau_m$ =20, 10 and 5, (a), (b) and (c), respectively.}
	\label{fig:figure14}	      
\end{figure}%
At a later stage, the density configuration evolves quite differently for inviscid HD fluid and GHD cases. In case of HD we observe two constantly rotating prominent crescent structures along with faint spirals. However, in the GHD cases, the crescent structures and the spirals get mixed due to the emerging TS wave (see Fig.~\ref{fig:figure15}). This result the absent of crescent structures and the whole structure evolves into spirals outward away from the center of rotation. As the coupling strength increases, the emission of TS wave becomes dominant. Comparison of last two snapshots from Fig.~\ref{fig:figure14}(a) ($\eta$=2.5; $\tau_m=20$), Fig.~\ref{fig:figure14}(b)  ($\eta$=2.5; $\tau_m=10$) and Fig.~\ref{fig:figure14}(c)  ($\eta$=2.5; $\tau_m=5$) clearly displays the aforementioned effects of the coupling strength on the evolution of the spirals. The presence of TS wave could be understood well by comparing vorticity contour plots for GHD mediums with different coupling strengths as shown in Fig.~\ref{fig:figure15}(a) ($\eta$=2.5; $\tau_m=20$), Fig.~\ref{fig:figure15}(b)  ($\eta$=2.5; $\tau_m=10$) and Fig.~\ref{fig:figure15}(c)  ($\eta$=2.5; $\tau_m=5$).   
\begin{figure}
	\includegraphics[width=1.0\textwidth]{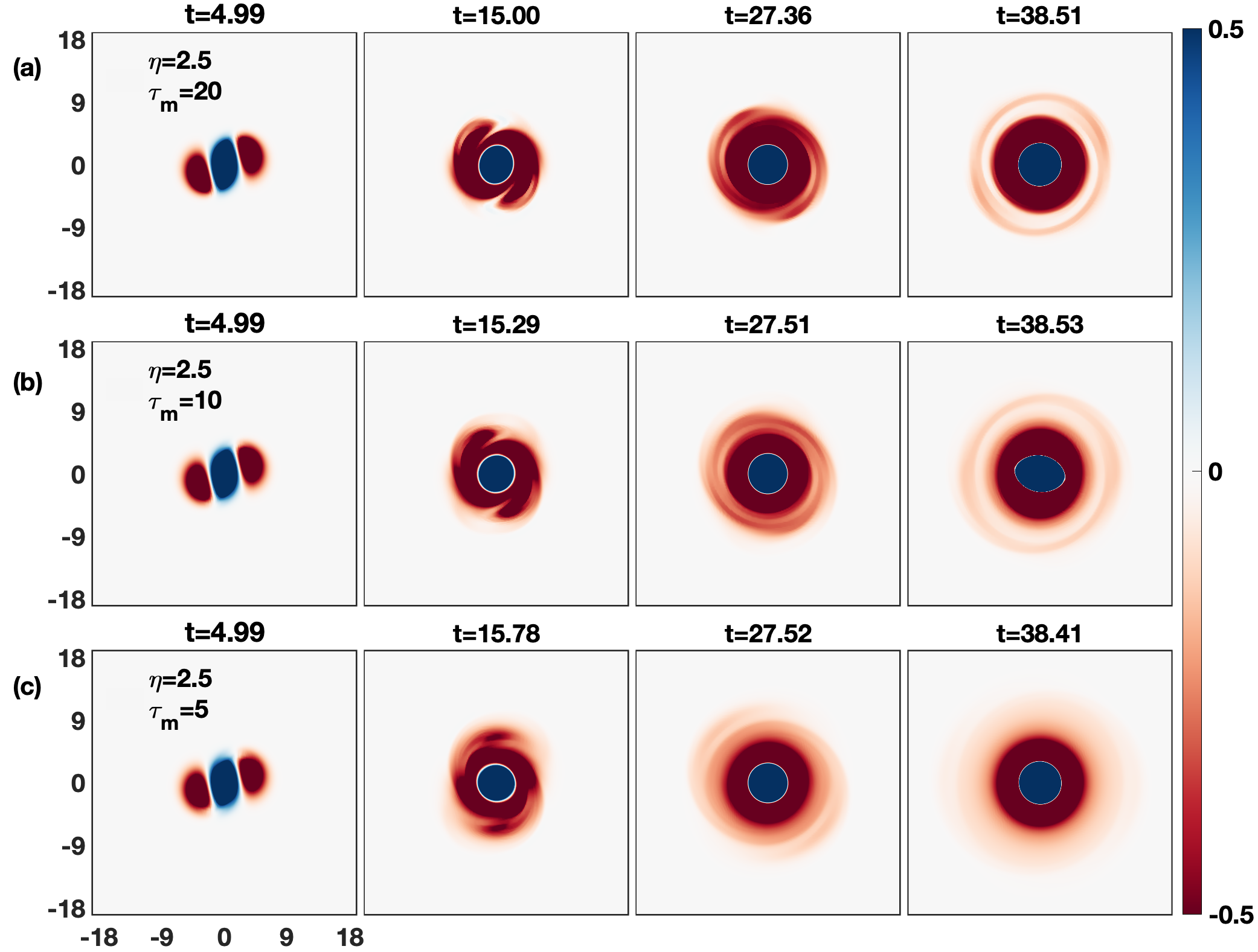}
\caption{Time evolution of bubble-droplet vorticity in viscoelastic fluids for fixed viscous term $\eta$=2.5; and varying coupling strength due to changing relaxation parameter $\tau_m$=20, 10 and 5, (a), (b) and (c), respectively.}
	\label{fig:figure15}	      
\end{figure}%
 Corresponding vorticity subplots for inviscid HD case in Fig.~\ref{fig:figure16} show absence of TS waves.    
\begin{figure}
	\includegraphics[width=1.0\textwidth]{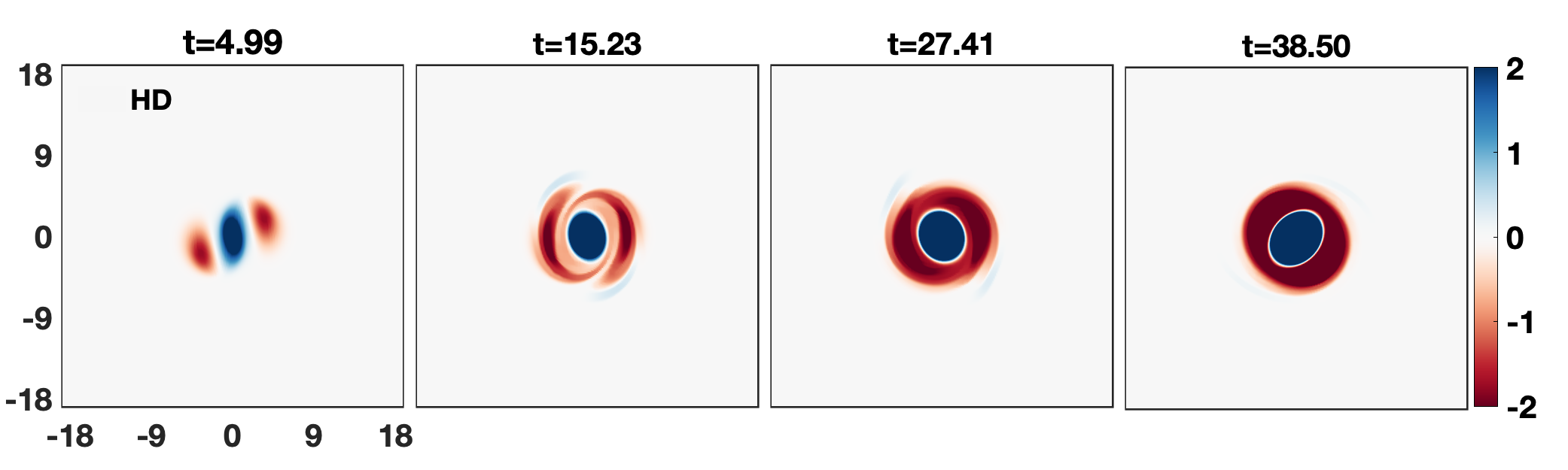}
	\caption{Time evolution of bubble-droplet vorticity for an inviscid hydrodynamic fluid corresponding to Fig.~\ref{fig:figure13}.}
	\label{fig:figure16}	      
\end{figure}%
Again, for the viscous fluid, due to the closeness Fig.~\ref{fig:figure17}  shows the higher rotation rate of spirals in comparison to the medium spaced (Fig.~\ref{fig:figure12}(a)) blobs.
\begin{figure}
	\includegraphics[width=1.0\textwidth]{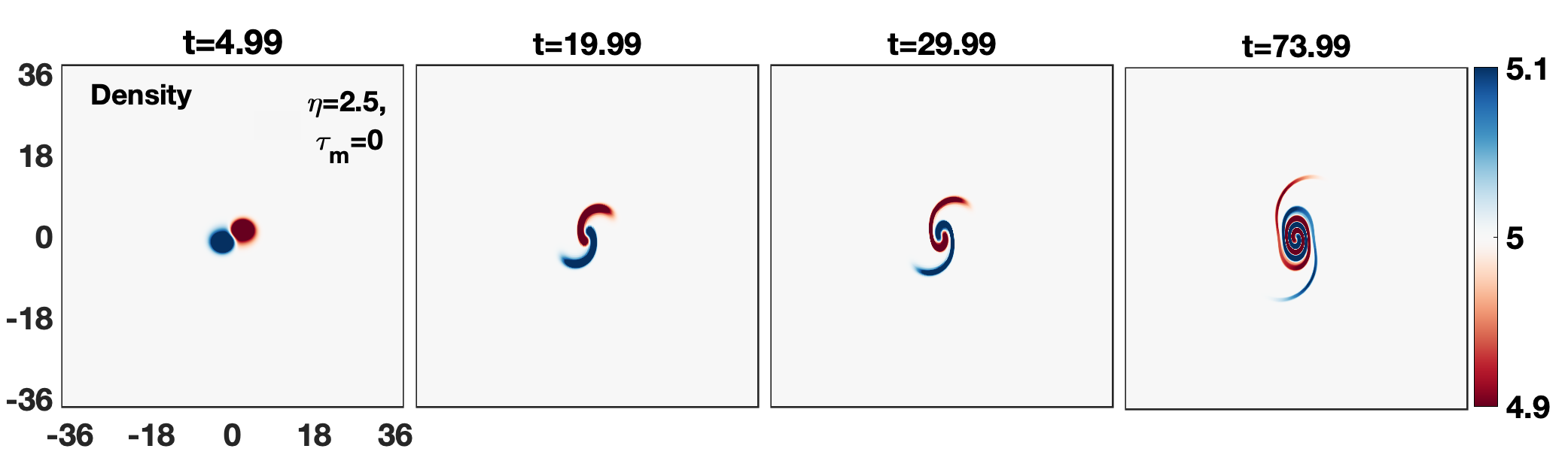}
				\caption{Time evolution of bubble-droplet density (top row) and vorticity (bottom row) for pure viscous HD fluid ($\eta$=2.5; $\tau_m$=0). The medium spaced blobs ($d{\approx}2a_c$) shows the higher rotaion rate of spirals than the medium spaced $d{>}2a_c$ (Fig.~\ref{fig:figure12}) case.}	
	\label{fig:figure17}	      
\end{figure}%

\subsection{Aligned vertically }
\begin{figure}
	\includegraphics[width=1.0\textwidth]{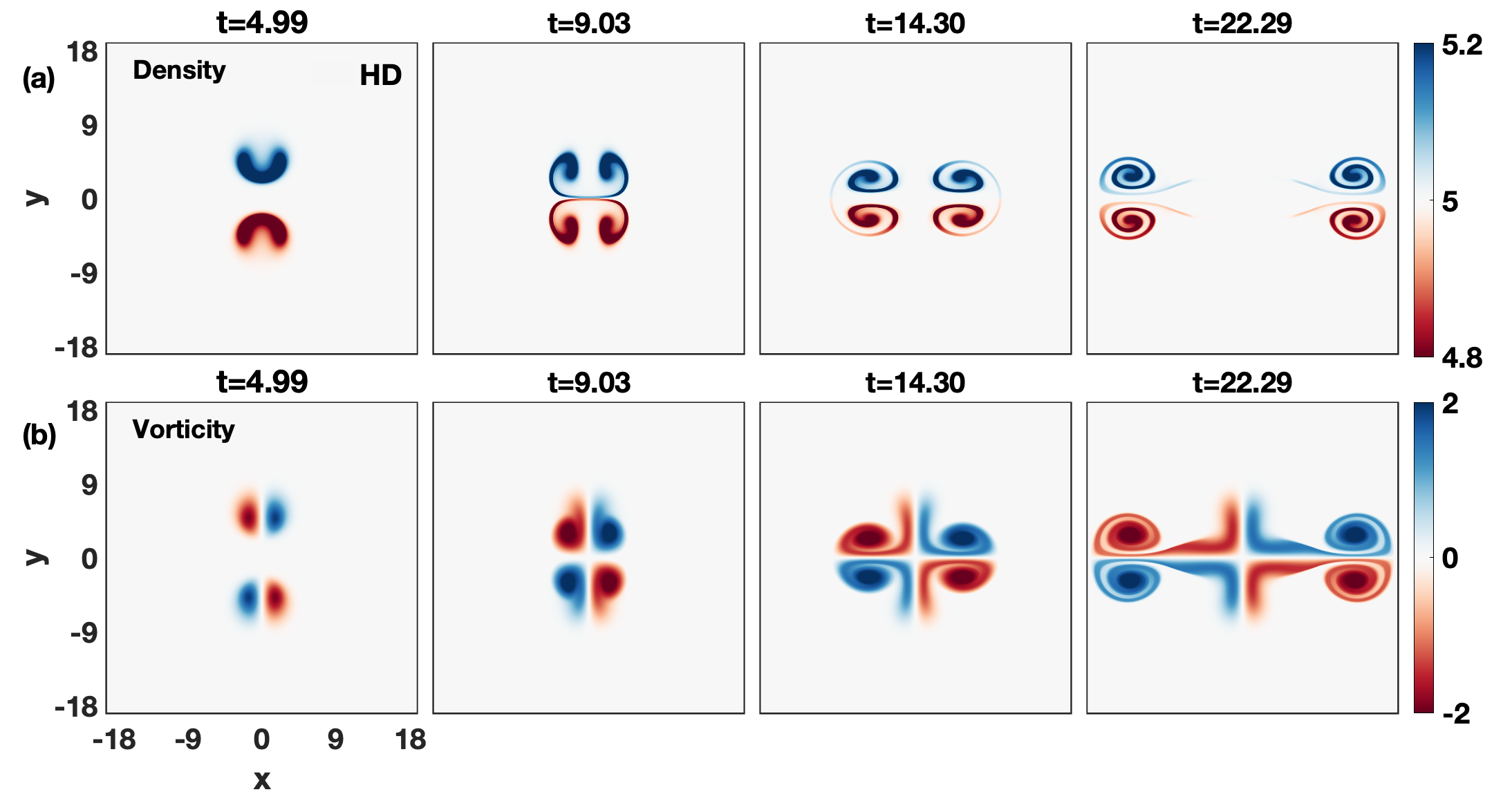}
	\caption{Time evolution of bubble-droplet density (top row) and vorticity (bottom row) for an inviscid hydrodynamic  fluid.  Both are  separated by distance d=12 units  $(d>>2a_c, a_c=2)$.}
	 \label{fig:figure18}
\end{figure}

\begin{figure}
	\includegraphics[width=1.0\textwidth]{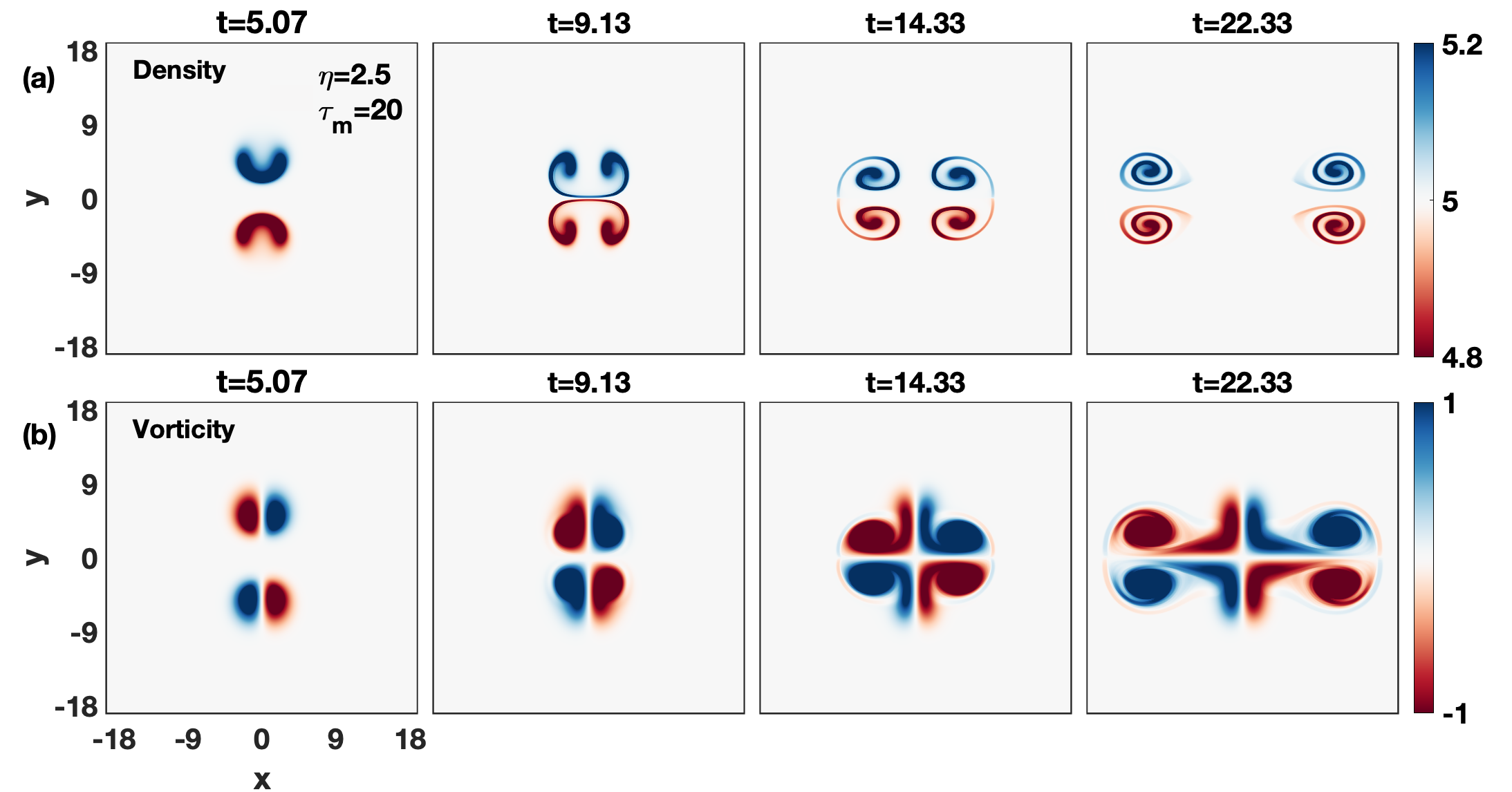}
	\caption{Time evolution of bubble-droplet density (top row) and vorticity (bottom row) for a viscoelatic fluid with $\eta=2.5$ and $\tau_m$=20.}
	\label{fig:figure19}	        
\end{figure}

\begin{figure}
	\includegraphics[width=1.0\textwidth]{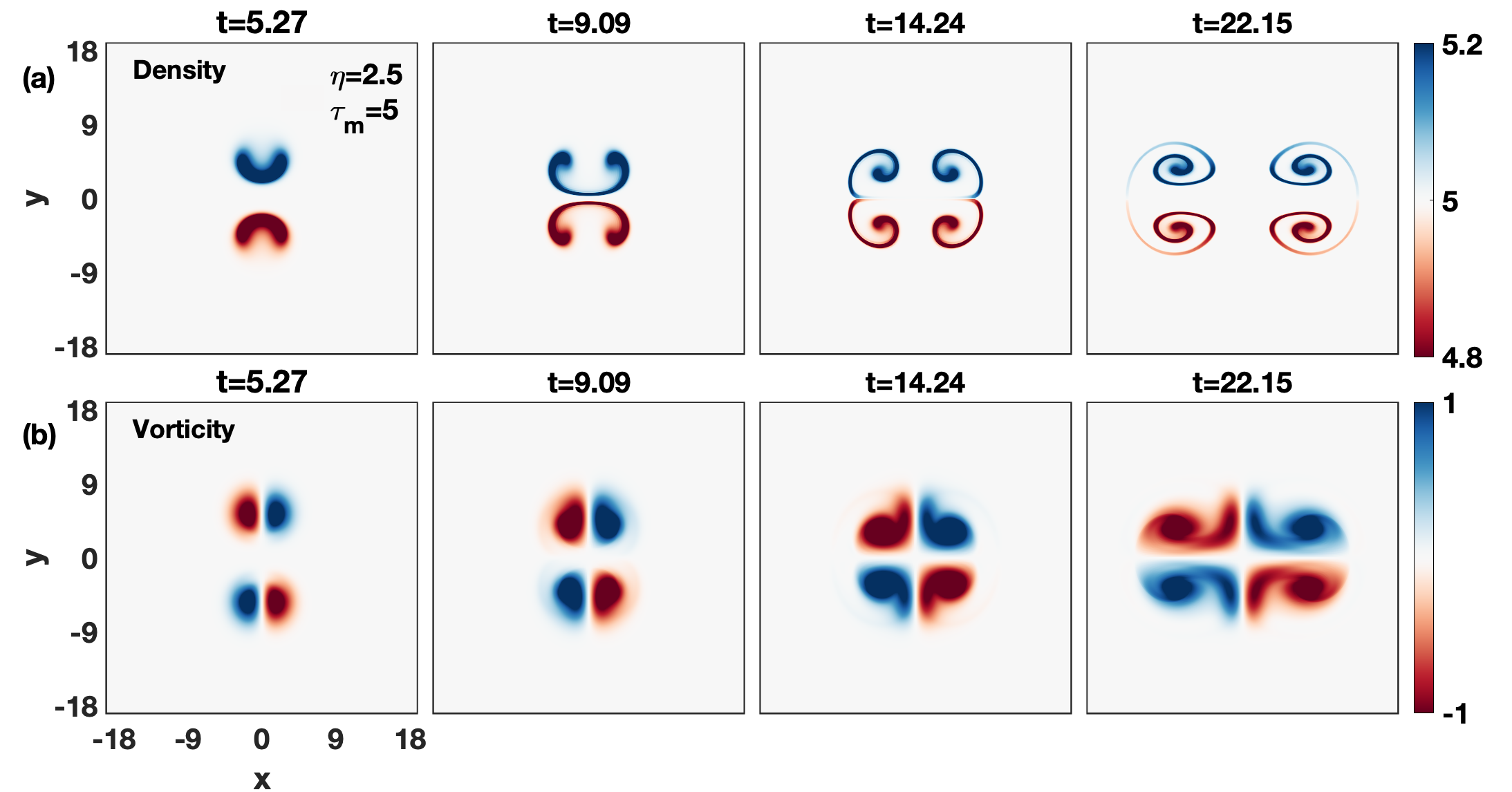}
	\caption{Time evolution of bubble-droplet density (top row) and vorticity (bottom row) for a viscoelatic fluid with $\eta=2.5$ and $\tau_m$=5.}
	\label{fig:figure20}	        
\end{figure}

We now consider the second arrangement (B) where the droplet is placed above bubble in a vertical column (see Fig.~\ref{fig:figure1}(b)). Here, coupling strength $({\eta}/{\tau_m})$ is the only varying parameter. We have a system of length $lx=ly=12{\pi}$ units with $512{\times}512$ grid points in both the x and y directions. The system along the x-axis and y-axis are from $-6{\pi}$ to  $6{\pi}$ units. For this case the values of the parameters $a_{c},~x_{c},~y_{c},~\rho^{\prime}$ for the droplet and bubble are 2.0, 0.0, $4\pi$, 0.5 and 2.0, 0.0, $-4\pi$, 0.5, respectively. For this configuration contrary to the previous case Fig.~\ref{fig:figure1}(a), there is no rotation of density blobs. Here, the falling droplet and rising bubble simply collide with each other during the course of evolution. Figure~\ref{fig:figure18}(a) displays the evolution of this density configuration for the hydrodynamic case. It is evident from the figure that as these two structures evolve, they hit each other and their blobs get separated. One blob from the bubble and one blob from the droplet pair with each other and move horizontally subsequently. 
\FloatBarrier
Figure~\ref{fig:figure19}(a) shows the evolution of same density configuration for GHD case ($\eta$=2.5; $\tau_m$=20). It is evident from figures that in comparison to the HD case, the horizontal propagation of the structures is slower and vertical separation between density blobs is larger with time. This horizontal reduction and vertical separation further get increase with increasing coupling strength shown in Fig.~\ref{fig:figure20}(a) ($\eta$=2.5; $\tau_m$=5). Such dynamics can be understood in terms of TS waves from the relative observations of Fig.~\ref{fig:figure19}(b) and Fig~\ref{fig:figure20}(b).
We know from the above discussion that  stronger coupling strength of a medium induces the faster TS waves. The faster TS waves result in a larger mutual pushing between these unlike-sign vorticity lobes in the vertical direction and gets separated vertically with time. Besides lobes separation, the emission of the TS wave reduced the strength of dipoles thereby reducing their propagation. The relative observations of Fig.~\ref{fig:figure20}(b) and Fig~\ref{fig:figure19}(b) clearly reflect the aforementioned fact.
\FloatBarrier

 Furthermore, the other possible dynamics by varying  initial spacing between the bubble and droplet can also be explored.
\section{Summary and conclusion}
\label{conclusions}
In part I of this paper, besides the Rayleigh-Taylor instability, we simulated individual dynamics of a rising bubble and a falling droplet in an incompressible limit of a strongly coupled dusty plasma (SCDP) using a generalized hydrodynamic viscoelastic fluid model. Here, in interest to understand the interactions between a rising bubble and a falling droplet, we extend this study to their combined evolution through two arrangements. A series of numerical simulations have been conducted.

In first arrangement the bubble and droplet are placed side-by-side in a row manually together at the same height for three different initial spacing: widely spaced $(d{\gg}{2a_{c}})$, medium spaced $(d{>}{2a_{c}})$, and closely spaced $(d{\approx}{2a_{c}})$, ${a_{c}}$ is core radius. In order to demonstrate how the appearance of  $\tau_m$ controls the viscous spreading of the vorticity lobes, for each case, the coupling strength has been introduced as the mild-strong ($\eta$=2.5, $\tau_m$=20), medium-strong ($\eta$=2.5, $\tau_m$=10) and strong or strongest ($\eta$=2.5, $\tau_m$=5), and pure viscosity ($\eta$=2.5, $\tau_m$=0). We have found that the net dynamic governs by the competition between the side-by-side attraction of two inner like-sign vorticity lobes due to the merging process and vertical motion of two dipolar vorticities due to the gravity. For the widely spaced bubble and droplet in media with the mild- and medium-strong coupling, the forward vertical motion of dipolar vorticities dominates over the side-by-side attraction. While for a strong coupling case the mutual attraction of vorticity lobes dominates which results in a rotating tripolar structure. For the medium spaced case, a rotating tripolar structure is found for all the three coupling parameters. The reduction in the size of these structures is observed with increasing coupling strength. While for an inviscid HD fluid the forward vertical motion of dipolar vorticities is observed. For the closely spaced case, tripolar structures appear for all the cases including the inviscid fluid. Second, the droplet is placed above the bubble in a vertical column with fixed initial spacing. Here, contrary to the previous case, there is no rotation of the bubble and droplet. Under the influence of gravity, the crescent-shaped bubble and droplet hit each other and their blobs get separated. One blob from the bubble and one blob from the droplet pair with each other and move horizontally subsequently. We observe their horizontal movement gets slower with increasing the coupling strength of the medium. 

In the next paper, it would also be interesting to see the homo-interactions between a pair of bubbles/droplets rising/falling side-by-side in homogeneous and heterogeneous background density media. In a heterogeneous medium, the radiating shear wave would play an important role because the shear waves move slower in the denser side and faster in the lighter side~\cite{dharodi2020rotating}.  

%
\bibliographystyle{unsrt}

\begin{thebibliography}{10}

\bibitem{dharodi2020numerical}
Vikram~S Dharodi and Amita Das.
\newblock A numerical study of gravity driven instability in strongly coupled
  dusty plasmas. part i: Rayleigh-taylor instability and buoyancy-driven
  instability.
\newblock {\em arXiv preprint arXiv:2006.12393}, 2020.

\bibitem{rayleigh1900investigation}
Lord Rayleigh.
\newblock Investigation of the character of the equilibrium of an
  incompressible heavy.
\newblock 1900.

\bibitem{taylor1950instability}
Geoffrey~Ingram Taylor.
\newblock The instability of liquid surfaces when accelerated in a direction
  perpendicular to their planes. i.
\newblock {\em Proceedings of the Royal Society of London. Series A.
  Mathematical and Physical Sciences}, 201(1065):192--196, 1950.

\bibitem{chandrhd}
S~Chandrasekhar.
\newblock Hydrodynamic and hydromagnetic stability, dover, new york, 1981.
\newblock 1981.

\bibitem{shew2006viscoelastic}
Woodrow~L Shew and Jean-Fran{\c{c}}ois Pinton.
\newblock Viscoelastic effects on the dynamics of a rising bubble.
\newblock {\em Journal of Statistical Mechanics: Theory and Experiment},
  2006(01):P01009, 2006.

\bibitem{gaudron2015bubble}
R~Gaudron, MT~Warnez, and E~Johnsen.
\newblock Bubble dynamics in a viscoelastic medium with nonlinear elasticity.
\newblock {\em Journal of Fluid Mechanics}, 766, 2015.

\bibitem{dollet2019bubble}
Benjamin Dollet, Philippe Marmottant, and Valeria Garbin.
\newblock Bubble dynamics in soft and biological matter.
\newblock {\em Annual Review of Fluid Mechanics}, 51:331--355, 2019.

\bibitem{dwyer1989calculations}
Harry~A Dwyer.
\newblock Calculations of droplet dynamics in high temperature environments.
\newblock {\em Progress in Energy and Combustion Science}, 15(2):131--158,
  1989.

\bibitem{cristini2004theory}
Vittorio Cristini and Yung-Chieh Tan.
\newblock Theory and numerical simulation of droplet dynamics in complex
  flows—a review.
\newblock {\em Lab on a Chip}, 4(4):257--264, 2004.

\bibitem{zhu2008three}
Xun Zhu, PC~Sui, and Ned Djilali.
\newblock Three-dimensional numerical simulations of water droplet dynamics in
  a pemfc gas channel.
\newblock {\em Journal of power sources}, 181(1):101--115, 2008.

\bibitem{leong2020droplet}
Fong~Yew Leong and Duc-Vinh Le.
\newblock Droplet dynamics on viscoelastic soft substrate: Toward coalescence
  control.
\newblock {\em Physics of Fluids}, 32(6):062102, 2020.

\bibitem{mokhtarzadeh1985dynamics}
MR~Mokhtarzadeh-Dehghan and AA~El-Shirbini.
\newblock Dynamics of two-phase bubble-droplets in immiscible liquids.
\newblock {\em W{\"a}rme-und Stoff{\"u}bertragung}, 19(1):53--59, 1985.

\bibitem{chen2008droplet}
Ruey-Hung Chen, David~S Tan, Kuo-Chi Lin, Louis~C Chow, Alison~R Griffin, and
  Daniel~P Rini.
\newblock Droplet and bubble dynamics in saturated fc-72 spray cooling on a
  smooth surface.
\newblock {\em Journal of heat transfer}, 130(10), 2008.

\bibitem{tabor2011homo}
Rico~F Tabor, Chu Wu, Hannah Lockie, Rogerio Manica, Derek~YC Chan, Franz
  Grieser, and Raymond~R Dagastine.
\newblock Homo-and hetero-interactions between air bubbles and oil droplets
  measured by atomic force microscopy.
\newblock {\em Soft Matter}, 7(19):8977--8983, 2011.

\bibitem{zhao2017droplet}
Lin Zhao, Michel~C Boufadel, Thomas King, Brian Robinson, Feng Gao, Scott~A
  Socolofsky, and Kenneth Lee.
\newblock Droplet and bubble formation of combined oil and gas releases in
  subsea blowouts.
\newblock {\em Marine Pollution Bulletin}, 120(1-2):203--216, 2017.

\bibitem{xie2017surface}
Lei Xie, Chen Shi, Xin Cui, and Hongbo Zeng.
\newblock Surface forces and interaction mechanisms of emulsion drops and gas
  bubbles in complex fluids.
\newblock {\em Langmuir}, 33(16):3911--3925, 2017.

\bibitem{van2001aeration}
George~A Van~Aken.
\newblock Aeration of emulsions by whipping.
\newblock {\em Colloids and Surfaces A: Physicochemical and Engineering
  Aspects}, 190(3):333--354, 2001.

\bibitem{liu2002fundamental}
J~Liu, T~Mak, Z~Zhou, and Z~Xu.
\newblock Fundamental study of reactive oily-bubble flotation.
\newblock {\em Minerals Engineering}, 15(9):667--676, 2002.

\bibitem{niewiadomski2007air}
Marcin Niewiadomski, Anh~V Nguyen, Jan Hupka, Jakub Nalaskowski, and Jan~D
  Miller.
\newblock Air bubble and oil droplet interactions in centrifugal fields during
  air-sparged hydrocyclone flotation.
\newblock {\em International journal of environment and pollution},
  30(2):313--331, 2007.

\bibitem{kong2019hydrodynamic}
Gaopan Kong, Haryo Mirsandi, KA~Buist, EAJF Peters, MW~Baltussen, and JAM
  Kuipers.
\newblock Hydrodynamic interaction of bubbles rising side-by-side in viscous
  liquids.
\newblock {\em Experiments in Fluids}, 60(10):155, 2019.

\bibitem{zhang2019vortex}
Jie Zhang, Long Chen, and Ming-Jiu Ni.
\newblock Vortex interactions between a pair of bubbles rising side by side in
  ordinary viscous liquids.
\newblock {\em Physical Review Fluids}, 4(4):043604, 2019.

\bibitem{Kaw_Sen_1998}
P.~K. Kaw and A.~Sen.
\newblock Low frequency modes in strongly coupled dusty plasmas.
\newblock {\em Physics of Plasmas}, 5(10):3552--3559, 1998.

\bibitem{Kaw_2001}
P.~K. Kaw.
\newblock Collective modes in a strongly coupled dusty plasma.
\newblock {\em Physics of Plasmas}, 8(5):1870--1878, 2001.

\bibitem{frenkel_kinetic}
J.~Frenkel.
\newblock {\em Kinetic Theory Of Liquids}.
\newblock Dover Publications, 1955.

\bibitem{dharodi2014visco}
V.~S. Dharodi, S.~K. Tiwari, and A.~Das.
\newblock Visco-elastic fluid simulations of coherent structures in strongly
  coupled dusty plasma medium.
\newblock {\em Physics of Plasmas}, 21(7):073705, 2014.

\bibitem{dharodi2016sub}
V.~S. Dharodi, A.~Das, B.~G. Patel, and P.~K. Kaw.
\newblock Sub-and super-luminar propagation of structures satisfying
  poynting-like theorem for incompressible generalized hydrodynamic fluid model
  depicting strongly coupled dusty plasma medium.
\newblock {\em Physics of Plasmas}, 23(1):013707, 2016.

\bibitem{boris_book}
J.~P. Boris, A.~M. Landsberg, E.~S. Oran, and J.~H. Gardner.
\newblock {\em LCPFCT A flux-corrected transport algorithm for solving
  generalized continuity equations}.
\newblock Technical Report NRL Memorandum Report 93-7192, Naval Research
  Laboratory, 1993.

\bibitem{swarztrauber1999fishpack}
P~Swarztrauber, R~Sweet, and John~C Adams.
\newblock Fishpack: Efficient fortran subprograms for the solution of elliptic
  partial differential equations.
\newblock {\em UCAR Publication, July}, 1999.

\bibitem{horstmann2014wake}
Jan~T Horstmann, Per Henningsson, Adrian~LR Thomas, and Richard~J Bomphrey.
\newblock Wake development behind paired wings with tip and root trailing
  vortices: consequences for animal flight force estimates.
\newblock {\em PloS one}, 9(3), 2014.

\bibitem{von2000vortex}
J~Von~Hardenberg, JC~McWilliams, A~Provenzale, A~Shchepetkin, and JB~Weiss.
\newblock Vortex merging in quasi-geostrophic flows.
\newblock {\em Journal of Fluid Mechanics}, 412:331--353, 2000.

\bibitem{meunier2005physics}
Patrice Meunier, St{\'e}phane Le~Diz{\`e}s, and Thomas Leweke.
\newblock Physics of vortex merging.
\newblock {\em Comptes Rendus Physique}, 6(4-5):431--450, 2005.

\bibitem{josserand2007merging}
Ch~Josserand and M~Rossi.
\newblock The merging of two co-rotating vortices: a numerical study.
\newblock {\em European Journal of Mechanics-B/Fluids}, 26(6):779--794, 2007.

\bibitem{kevlahan1997vorticity}
NK-R Kevlahan and Marie Farge.
\newblock Vorticity filaments in two-dimensional turbulence: creation,
  stability and effect.
\newblock {\em Journal of Fluid Mechanics}, 346:49--76, 1997.

\bibitem{dharodi2020rotating}
Vikram~S Dharodi.
\newblock Rotating vortices in two-dimensional inhomogeneous strongly coupled
  dusty plasmas: Shear and spiral density waves.
\newblock {\em Physical Review E}, 102(4):043216, 2020.

\bibitem{van1991formation}
GJF van Heijst, RC~Kloosterziel, and CWM Williams.
\newblock Formation of a tripolar vortex in a rotating fluid.
\newblock {\em Physics of Fluids A: Fluid Dynamics}, 3(9):2033--2033, 1991.

\bibitem{huang2005physical}
Mei-Jiau Huang.
\newblock The physical mechanism of symmetric vortex merger: A new viewpoint.
\newblock {\em Physics of Fluids}, 17(7):074105, 2005.

\end{thebibliography}

\end{document}